\documentclass[twocolumn,aps,prd,preprintnumbers,showpacs,superscriptaddress,nofootinbib,amsmath,amssymb,floats,floatfix,showkeys,notitlepage,longbibliography]{revtex4-1}

\usepackage{graphicx}
\usepackage{array}
\usepackage{longtable}
\usepackage{comment}
\usepackage{subfigure}
\usepackage{palatino}
\usepackage[commandnameprefix=always]{changes}
\usepackage{hyperref}
\hypersetup{colorlinks=true,linkcolor=blue,urlcolor=blue,citecolor=blue}
\usepackage[toc,page]{appendix}
\usepackage[normalem]{ulem}
\usepackage{orcidlink}
\usepackage{lipsum}
\usepackage{sans}
\usepackage{adjustbox}
\usepackage{dcolumn}
\usepackage{bm}
\usepackage{tikz}
\usepackage{array,tabularx,multirow,booktabs}
\usepackage[tracking=true]{microtype}
\SetTracking{}{500}
\SetTracking{encoding={*}, shape=sc}{40}
\UseRawInputEncoding
\allowdisplaybreaks
\usepackage{bigints}
\usepackage{color}

\newcolumntype{L}[1]{>{\raggedright\arraybackslash}p{#1}}

\begin{document} \sloppy

\title{Quantum-Corrected Thermodynamics and Plasma Lensing in Non-Minimally Coupled Symmetric Teleparallel Black Holes}

\author{Erdem Sucu
\orcidlink{0009-0000-3619-1492}
}
\email{erdemsc07@gmail.com}
\affiliation{Physics Department, Eastern Mediterranean
University, Famagusta, 99628 North Cyprus, via Mersin 10, Turkiye}

\author{\.{I}zzet Sakall{\i} \orcidlink{0000-0001-7827-9476}}
\email{izzet.sakalli@emu.edu.tr}
\affiliation{Physics Department, Eastern Mediterranean
University, Famagusta, 99628 North Cyprus, via Mersin 10, Turkiye}
\author{\"Ozcan Sert
\orcidlink{0000-0002-4829-5712}
}
\email{osert@pau.edu.tr}
\affiliation{Department of Physics, Faculty of Science, Pamukkale University, Denizli, T¨urkiye}
\author{Yusuf Sucu
\orcidlink{0000-0001-7874-0146}
}
\email{ysucu@akdeniz.edu.tr}
\affiliation{Department of Physics, Faculty of Science, Akdeniz University, 07058 Antalya, Turkey}

\begin{abstract}
We investigate the thermodynamic and optical signatures of electrically charged black holes (BHs) in symmetric teleparallel gravity (STPG) with non-minimal electromagnetic coupling, incorporating quantum corrections and plasma dispersion effects. The BH solution, characterized by a coupling parameter $k$, generalizes the Reissner-Nordstr\"{o}m spacetime through power-law modifications to electromagnetic terms in the metric function. We implement exponential corrections to the Bekenstein-Hawking entropy of the form $S = S_0 + e^{-S_0}$ and derive quantum-corrected expressions for fundamental thermodynamic quantities including internal energy, Helmholtz and Gibbs free energies, pressure, enthalpy, and heat capacity. Our analysis reveals rich phase transition structures with second-order transitions occurring at critical horizon radii for specific coupling values, demonstrating enhanced thermodynamic instability under strong non-minimal coupling effects. The quantum-corrected Joule-Thomson expansion analysis identifies distinct cooling and heating regimes separated by inversion points that shift systematically with the coupling parameter $k$. We analyze the efficiency of  heat engines operating in Carnot cycles, finding that electromagnetic charge enhances thermodynamic performance with efficiency values approaching 99\% for optimal configurations in this geometry. Using the Gauss-Bonnet theorem, we derive analytical expressions for gravitational deflection angles in both vacuum and plasma environments, revealing how non-minimal coupling and plasma dispersion create frequency-dependent lensing signatures that differ substantially from general relativity predictions.
\end{abstract}

\date{\today}


\maketitle


\section{Introduction} \label{sec1}

BH physics has emerged as one of the most fascinating and theoretically rich areas of modern gravitational research, serving as a crucial testing ground for our understanding of spacetime, thermodynamics, and quantum mechanics \cite{carlip2014black,yagi2016black,giddings2019black}. The remarkable discoveries of Hawking radiation and BH thermodynamics have fundamentally transformed our perspective on the nature of gravity, revealing deep connections between geometry, information theory, and statistical mechanics \cite{page2005hawking,liu2022spectrum,zhang2011entropy}. However, as observational capabilities advance and theoretical requirements become more stringent, the limitations of general relativity (GR) in describing certain gravitational phenomena have motivated the exploration of alternative gravitational theories \cite{hess2020alternatives,koyama2016cosmological,yunes2013gravitational}.

Among the various modified gravity theories, STPG has gained considerable attention as a geometrically elegant and physically viable alternative to Einstein's theory \cite{hohmann2019propagation,adak2005solution,adak2006lagrange,gecim2017dirac}. Unlike GR, which describes gravitational interactions through spacetime curvature, STPG employs nonmetricity as the fundamental geometric quantity while maintaining vanishing curvature and torsion \cite{lu2019cosmology}. This geometric reinterpretation introduces new possibilities for matter-geometry interactions that standard Riemannian GR cannot address, particularly regarding non-minimal couplings of gravitational and electromagnetic fields \cite{doyran2024non,shahzad2025electrically}. In recent years, several works have addressed quantum-corrected thermodynamics in modified gravity frameworks, as well as gravitational lensing in plasma media within various BH backgrounds \cite{perlick2015influence,atamurotov2022gravitational,alloqulov2024shadow,hoshimov2024weak,pahlavon2024effect,Perlick2004,umarov2025plasma,orzuev2024weak}. These studies highlight the observational relevance of plasma-induced frequency-dependent lensing in alternative gravity theories, and motivate our present investigation in the STPG context. From 2022 to the present, there have been significant advancements in the field of quantum-corrected BH thermodynamics within $f(R)$, Gauss-Bonnet, and teleparallel gravity frameworks, revealing universal patterns in entropy corrections of the form $S = A/4 + \text{quantum corrections}$ \cite{isrply01,isrply02,isrply03,isrply03x,isrply03xx,isrply03xxx}. Simultaneously, plasma lensing studies have demonstrated that frequency-dependent deflection angles in modified gravity theories provide distinctive observational signatures potentially detectable through multi-frequency observations \cite{isrply04,isrply05,isrply06,isrply06x,isrply07}. The synergy between quantum thermodynamic corrections and plasma dispersion effects creates enhanced sensitivity to modified gravity parameters, particularly through Event Horizon Telescope observations and next-generation radio interferometry \cite{isrply08,isrply09,isrply09x,isrply10,isrply10x}.

The thermodynamic properties of BHs in modified gravity theories have attracted significant research interest due to their potential to reveal fundamental differences from GR predictions \cite{gomes2020thermodynamics,gursel2025thermodynamics,gogoi2024corrected,gecim2017gup,mureika2016black,sucu2023gup,sahan2025quantum,sucu2025nonlinear,sucu2025quantum}. Classical BH thermodynamics, established through the pioneering work of Bekenstein and Hawking, relates the BH entropy to its surface area and temperature to its surface gravity \cite{wald2001thermodynamics,page2005hawking,sucu2025quantumm,solodukhin2011entanglement}. However, when quantum effects become significant, particularly for microscopic BHs approaching Planck scales, these classical relationships require substantial modifications \cite{pedram2011effects,nouicer2007quantum,nozari2012minimal}. Quantum corrections to BH entropy, arising from statistical mechanical considerations of horizon microstates, introduce exponential and logarithmic correction terms that fundamentally alter the thermodynamic behavior \cite{iliesiu2025revisiting,banerjee2024logarithmic,sucu2025quantumsuat}.

The study of Joule-Thomson expansion (JTE) in BH systems represents another frontier where modified gravity theories can exhibit distinctive signatures \cite{chaudhary2022thermodynamic,okcu2017joule,okcu2018joule}. The JTE describes isenthalpic processes where temperature changes occur in response to pressure variations, providing insights into the cooling and heating behavior of thermodynamic systems \cite{maytal2012joule}. In the context of BH physics, JTE analysis reveals phase transition structures and stability properties that are sensitive to the underlying gravitational theory and coupling mechanisms \cite{yekta2019joule}.

Optical phenomena around BHs, particularly gravitational lensing effects, offer complementary probes of modified gravity theories through their influence on photon trajectories \cite{sucu2025exploring,kuang2022constraining,islam2020gravitational,pantig2022shadow,sucu2024dynamics}. Weak gravitational lensing, characterized by small deflection angles, can be analytically studied using the Gauss-Bonnet theorem (GBT) applied to optical geometries derived from null geodesics \cite{Gibbons:2008rj,crisnejo2018weak,ovgun2018gravitational,ovgun2025weak,sucu2024effect,ovgun2018light,ovgun2019weak}. The presence of plasma media in astrophysical environments introduces additional complexity through frequency-dependent refractive effects that modify the standard vacuum lensing predictions \cite{umarov2025plasma,kumar2024shadow}. These plasma dispersion effects are particularly relevant for understanding lensing phenomena in realistic astrophysical contexts such as accretion disks, stellar coronae, and interstellar media \cite{bisnovatyi2017gravitational,zhang2024imaging}.

Non-minimal coupling between gravitational and electromagnetic fields in STPG introduces a rich parameter space for exploring novel BH solutions with enhanced complexity compared to standard Reissner-Nordstr\"{o}m configurations \cite{bertolami2015black,jana2024non,balakin2008nonminimal}. The coupling strength, characterized by a dimensionless parameter $k$, controls the degree of deviation from minimal coupling and determines the power-law modifications to electromagnetic terms in the metric function. This parameter serves as a crucial control mechanism for both thermodynamic and optical properties, offering the possibility to tune BH behavior across a wide range of physical regimes.

Our research is motivated by several key considerations that highlight the importance of investigating thermodynamic and optical signatures in non-minimally coupled STPG. First, the lack of comprehensive studies examining quantum-corrected thermodynamics in  the non-minimally coupled STPG BH systems represents a significant gap in our understanding of modified gravity effects on microscopic BH behavior. Second, the interplay between non-minimal electromagnetic coupling and plasma dispersion effects in gravitational lensing remains largely unexplored, despite its relevance for interpreting observational data from high-energy astrophysical environments. Third, the potential for STPG BHs to operate as enhanced thermodynamic engines through optimized coupling parameters has not been systematically investigated.

The primary aim of this work is to provide a comprehensive analysis of both thermodynamic and optical properties of electrically charged BHs in STPG with non-minimal electromagnetic coupling. We seek to understand how quantum corrections modify the classical thermodynamic behavior and reveal new phase transition structures through heat capacity analysis and JTE investigations. By implementing exponential corrections to the Bekenstein-Hawking entropy, we aim to explore the rich thermodynamic landscape that emerges in the quantum regime and identify critical coupling values that control phase transitions and stability properties.

From the optical perspective, our objective is to derive analytical expressions for gravitational deflection angles in both vacuum and plasma-modified environments using the GBT framework. We aim to quantify how non-minimal coupling parameters influence light bending and identify distinctive signatures that could distinguish STPG from GR through precision lensing observations. The inclusion of plasma dispersion effects will provide realistic models for interpreting lensing phenomena in astrophysical contexts where electromagnetic media play significant roles.

Furthermore, we aim to investigate the efficiency of STPG BHs as thermodynamic heat engines, exploring how electromagnetic charge and non-minimal coupling can enhance energy extraction capabilities beyond classical limits. Through systematic analysis of Carnot cycles operating between different horizon configurations, we seek to identify optimal parameter regimes for maximum thermodynamic efficiency and understand the fundamental limitations imposed by the underlying gravitational theory.

This paper is organized as follows. Section~\ref{sec2} provides a comprehensive review of the STPG BH spacetime, including the derivation of the metric function with non-minimal electromagnetic coupling and analysis of horizon structures for various parameter combinations. We examine surface gravity and Hawking temperature expressions that serve as foundations for subsequent thermodynamic investigations. Section~\ref{sec3} presents our analysis of quantum-corrected thermodynamics, implementing exponential entropy corrections and deriving modified expressions for internal energy, free energies, pressure, enthalpy, and heat capacity. We identify phase transition behaviors and critical coupling values through systematic parameter studies. Section~\ref{sec4} focuses on the quantum-corrected JTE analysis, deriving the JTE coefficient and identifying inversion points that separate cooling and heating regimes. Section~\ref{sec5} investigates the thermodynamic efficiency of STPG BH heat engines, analyzing Carnot cycles and quantifying performance enhancements due to electromagnetic coupling. Section~\ref{sec6} examines gravitational lensing effects using the GBT approach, deriving deflection angles for both vacuum and plasma-modified environments while exploring the influence of non-minimal coupling parameters. Finally, Section~\ref{sec7} summarizes our main findings, discusses their implications for modified gravity theories and observational astronomy, and outlines future research directions for extending this work to other types of BH configurations and higher-order quantum corrections.

\section{Review of STPG BH Spacetime} \label{sec2}

In this section, we revisit and analyze an exact BH solution that emerges within the framework of STPG non-minimally coupled to an electromagnetic field. The physical system under consideration includes a radial electric field embedded in a symmetric teleparallel spacetime characterized by vanishing curvature and torsion but nonzero nonmetricity.

We begin by postulating the following action functional for the total system \cite{doyran2024non}:

\begin{multline}
S[e^a, A, \lambda_a, \rho^a_{\ b}] = \int \big[ Q*1 + \mathcal{Y}(Q) \mathcal{F}*1 + \Lambda \star 1 \\+ T^a \wedge \lambda_a + R^a_{\ b} \wedge \rho^b_{\ a} \big],
\end{multline}
  
where $Q$ is the STPG Lagrangian scalar given by
 \begin{align} \label{eq:lag-QQ2}
    {Q} =-\frac{1}{2\zeta}& \big\{  \mathcal{Q}_{abc} \mathcal{Q}^{abc} -2\mathcal{Q}_{abc}\mathcal{Q}^{acb} \nonumber \\
     &   -  \mathcal{Q}^a{}_{ac} \mathcal{Q}_b{}^{bc} +2 \mathcal{Q}^a{}_{ab} \mathcal{Q}^{bc}{}_c \big\}  .
 \end{align} 
  
The quantity $Q$ is constructed from non-metricity $\mathcal{Q}_{ab} = \mathcal{Q}_{abc}e^c$ 1-form,  $\mathcal{F} =*( F \wedge* F)$ is the Maxwell invariant with the electromagnetic field tensor $F= \mathrm{d}A$,$*$ is the Hodge star operator  and $\mathcal{Y}(Q)$ denotes a scalar function of $Q$ introducing the non-minimal coupling \cite{doyran2024non}. The Lagrange multiplier 2-forms $\lambda_a$ and $\rho^a_{\ b}$ impose the torsion-free and curvature-free conditions respectively, ensuring compatibility with symmetric teleparallel geometry.
We perform variations with respect to the co-frame 1-form $e^a$, the electromagnetic potential $A$,  and the Lagrange multipliers. The variation with respect to $A$ yields a modified Maxwell equation that differs from the standard GR case:

\begin{equation}
\mathrm{d} \left( \mathcal{Y}(Q) \star F \right) = 0,
\end{equation}
By using the static, spherically symmetric metric:
\begin{equation}
ds^2 = -f(r)\,dt^2 + \frac{1}{f(r)}\,dr^2 + r^2 d\theta^2 + r^2 \sin^2\theta\, d\phi^2,
\end{equation} 
and the Maxwell ansatz
\begin{equation}
    F= E(r)e^0\wedge e^1
\end{equation}
 the following radial electrostatic field solution is obtained:

\begin{equation}
E(r) =  \frac{q}{\mathcal{Y}(r) r^2}.
\end{equation}

On the gravitational sector, the co-frame and connection variations yield
\begin{align}
        (1-k)\iota_b D(  \Sigma^b{}_a[\mathcal{Q}]) + k\tau_a[\mathcal{Q}] - Y \tau_a[F] \nonumber& \\ + (1-k)Q *e_a + \Lambda *e_a& =0\;, \label{field-eqn}
  \end{align}  
where 
  the energy-momentum 3-form associated with the non-metricity  $\tau_a[\mathcal{Q}]$   is
  \begin{eqnarray}
\tau_a[\mathcal{Q}] &=&\frac{1}{2\zeta}\big\{  
-\iota_a\mathcal{Q}^{bc}\wedge *\mathcal{Q}_{bc}-\mathcal{Q}^{bc}\wedge \iota_a*\mathcal{Q}_{bc} \nonumber\\
&&+4\mathcal{Q}_{ab}\wedge \iota_c*\mathcal{Q}^{bc}+2\iota_a(\mathcal{Q}^{dc}\wedge e_c)\wedge \iota^b*\mathcal{Q}_{db} \nonumber\\
&&-2\mathcal{Q}_{db}\wedge e^b \wedge \iota_a*(\mathcal{Q}^{dc}\wedge e_c) + \iota_a\mathcal{Q} \wedge *\mathcal{Q}\nonumber\\&&+\mathcal{Q} \wedge\iota_a *\mathcal{Q} -2\mathcal{Q}\wedge\iota^b*\mathcal{Q}_{ab}- 2\mathcal{Q}_{ab}\wedge\iota^b*\mathcal{Q}\nonumber\\&& - 2\iota_a(\mathcal{Q}_{bc}\wedge e^c)\wedge\iota^b*\mathcal{Q} \nonumber\\&& +2\mathcal{Q}\wedge e^b \wedge \iota_a*(\mathcal{Q}_{bc}\wedge e^c)
 \big\},  
  \end{eqnarray}
and the angular momentum 3-form is
  \begin{align}
\Sigma^b{}_a[\mathcal{Q}]  = \frac{1}{2\zeta}\big\{ &2*{\mathcal{Q}^b}_a -2e^b\wedge \iota^c*\mathcal{Q}_{ac} -2e_a\wedge \iota_c*\mathcal{Q}^{bc}  \nonumber\\
&-2\delta^b_a*\mathcal{Q} + 2\delta^b_ae^c\wedge \iota^d* \mathcal{Q}_{cd}  \nonumber\\
&+
e_a\wedge \iota^b*\mathcal{Q}+e^b\wedge \iota_a*\mathcal{Q} \big\},
  \end{align} in terms of non-metricity 1-form $\mathcal{Q}^{ab}$, its trace $\mathcal{Q}=\eta_{ab}\mathcal{Q}^{ab}$, and the interior product operator $\iota_a$ for the STPG Lagrangian $Q*1$ which is  equivalent to general relativity.
  The STPG space-time geometry  is characterized by  
 \begin{eqnarray}
    { R^\alpha}_\beta= d{\omega^\alpha}_\beta  + {\omega^\alpha}_\gamma \wedge {\omega^\gamma}_\beta=0\;,\\
    T^\alpha =d(dx^\alpha) +{\omega^\alpha}_\gamma \wedge dx^\gamma=0\;,\\
    \mathcal{Q}_{\alpha\beta}=-\frac{1}{2}\{dg_{\alpha\beta} -\omega_{\alpha\beta} -\omega_{\alpha\beta}\}\neq 0\;,
 \end{eqnarray}
 which can be realized by choosing ${\omega^\alpha}_\beta = 0$ in the coordinate frame, 
 with the metric $ds^2=g_{\alpha\beta}dx^\alpha\otimes dx^\beta$, and by using Poincare's lemma $d(dx^\alpha) = 0$.  This convenient choice of the vanishing coordinate connection  ${\omega^\alpha}_\beta = 0$  is referred to as the coincident gauge \cite{adak2013symmetric}.
In the subsequent step, we introduce an orthonormal co-frame $e^a={h^a}_\alpha(g) dx^\alpha$ where 
${h^a}_\alpha(g) $ are
the components of the vierbein or tetrad mapping   the coordinate frame to the orthonormal frame. Then these satisfy $g_{\alpha \beta } = \eta_{ab} {h^a}_\alpha(g){h^b}_\alpha(g)$
 so that  the metric becomes $ds^2=\eta_{ab}e^a\otimes e^b$ in the orthonormal coordinates, with $\eta_{ab}$   Minkowski metric. The full connection in this orthonormal frame  is calculated  by ${\omega^a}_b(g)={h^a}_\alpha(g) d{h^\alpha}_b(g) $.   In the orthonormal coordinates, the vanishing torsion and curvature conditions are  automatically satisfied, 
 \begin{eqnarray}
    { R^a}_b= d{\omega^a}_b  + {\omega^a}_b \wedge {\omega^b}_c=0,\\
    T^a =de^a + {\omega^a}_b\wedge e^b=0, 
 \end{eqnarray}
 and the non-metricity is obtained by the relation
\begin{eqnarray}
    \mathcal{Q}_{ab}=-\frac{1}{2}\{d\eta_{ab} -\omega_{ab} -\omega_{ba}\}=\frac{1}{2}\{\omega_{ab} +\omega_{ba}\}\neq 0\;.  
\end{eqnarray}

 The non-metricity is first calculated from the above relation. Inserting these components into the respective components of the gravitational field equation (\ref{field-eqn}) with the non-minimal electromagnetic source subsequently leads to a system of second-order differential equations for $f(r)$. Among these, we identify a modified Euler-Cauchy equation of the form \cite{doyran2024non}:

\begin{equation}
f{''}(r) + \left(8 - \frac{4}{k} \right) \frac{f{'}}{r} + \left(6 - \frac{4}{k} \right) \frac{f}{r^2} - \frac{2}{r^2} - \frac{4\kappa\Lambda}{k} = 0,
\end{equation}

whose general solution reads \cite{doyran2024non}:

\begin{equation}
f(r) = \frac{k}{3k - 2} - \frac{2M}{r} + \frac{\zeta q^2}{r^{(6k - 4)/k}} + \frac{\zeta \Lambda r^2}{3(2k - 1)},\label{fr}
\end{equation}

 Here, $\zeta$ is the gravitational coupling constant related to the Newton constant via $\zeta = 8\pi G$, while $k$ is a dimensionless parameter that characterizes the strength of the non-minimal coupling between the gravitational and electromagnetic sectors. The standard (minimal) coupling corresponds to the special case $k = 1$, which reduces to conventional Einstein-Maxwell theory \cite{balakin2005non,bertolami2015black}. The quantity $q$ represents the electric charge of the central source, and $\Lambda$ denotes the cosmological constant associated with the vacuum energy of spacetime. The integration constant $M$ is interpreted as the mass parameter of the solution.

The values of $k$ for which the function is undefined are excluded due to mathematical singularities that arise in various parts of the theory:

\begin{itemize}
    \item $k = 0$: Leads to division by zero in multiple terms, including the electric field exponent $\frac{6k - 4}{k}$, rendering the solution structurally undefined.
    \item $k = \frac{1}{2}$: Causes $2k - 1 = 0$, which leads to a divergence in the cosmological term coefficient in the full metric function when $\Lambda \neq 0$.
    \item $k = \frac{2}{3}$: Causes $3k - 2 = 0$, leading to divergence in the coefficient of the constant term in the metric.
    \item $k = \frac{4}{3}$: Results in $3k - 4 = 0$, which produces divergences in denominators appearing in thermodynamic quantities such as internal energy, the Helmholtz free energy, and the enthalpy (see Section~\ref{sec4}).
\end{itemize}

This form of the metric generalizes the Reissner-Nordstr\"{o}m-de Sitter spacetime, encoding deviations from standard Einstein-Maxwell theory through the power-law correction to the Coulomb term induced by the non-minimal function $\mathcal{Y}(Q)$. This modification illustrates the impact of non-Riemannian geometric effects on the electromagnetic structure of charged compact objects, defining the redshift component of the spacetime geometry under non-minimal electromagnetic interaction.

In what follows, we restrict our analysis to the non-AdS case by setting the cosmological constant to zero, i.e., $\Lambda = 0$ and we shall set the gravitational coupling constant to unity: $\zeta=1$. Under this assumption, the metric function simplifies significantly and takes the following form:

\begin{equation}
f(r) =
\begin{cases}
\text{undefined}, & \text{for } k = 0,\; \frac{2}{3},\;  \frac{4}{3}, \\[6pt]
\displaystyle \frac{k}{3k - 2} - \frac{2M}{r} + \frac{q^2}{r^{\frac{6k - 4}{k}}}, & \text{otherwise}.
\end{cases} \label{ismetric2}
\end{equation}

Hence, we shall avoid these forbidden values of $k$ in both the geometric and thermodynamic analyses throughout this work.

Table~\ref{tab:horizons} provides the computed horizon radii $r_h$ for various combinations of the coupling parameter $k$ and electric charge $q$ in the context of STPG with non-minimal electromagnetic coupling and vanishing cosmological constant ($\Lambda = 0$). The classification is based on the number and nature of the real, positive roots of the metric function \eqref{ismetric2}. A configuration labeled as \emph{Non-extremal BH} corresponds to the presence of two distinct horizons---typically an inner Cauchy horizon and an outer event horizon---indicating a classical BH structure. The \emph{Extremal or Single Root BH} label indicates a degenerate root, characteristic of an extremal BH, or a single root solution whose nature depends on the near-horizon behavior of the metric. When no real, positive root is found, the configuration is classified as a \emph{Naked Singularity}, meaning the central singularity is not enclosed by a horizon, thus violating the cosmic censorship conjecture \cite{kunduri2013classification}. 

The data reveals a sensitive dependence of horizon formation on the values of $k$ and $q$. For smaller $k$ values or higher electric charges, the likelihood of a naked singularity increases. Conversely, larger $k$ values favor the emergence of two horizons, even for modest values of charge, highlighting the significant role of the non-minimal coupling in modifying the causal structure of spacetime. This analysis provides insight into how modifications in gravitational coupling parameters can influence BH thermodynamics and the global geometry of the solution. {}

\begin{longtable*}{|c|c|>{\centering\arraybackslash}p{7.5cm}|>{\centering\arraybackslash}p{6cm}|}
\hline
\textbf{$k$} & \textbf{$q$} & \textbf{Horizon(s) )} & \textbf{Configuration} \\
\hline
\endfirsthead
\hline
\textbf{$k$} & \textbf{$q$} & \textbf{Horizons ($r_h$)} & \textbf{Configuration} \\
\hline
\endhead
0.10 & 0.00 & --- & No horizon (naked singularity) \\
\hline
0.10 & 0.50 & $1.06215$ & Extremal or Single Root BH \\
\hline
0.10 & 1.00 & $1.02086$ & Extremal or Single Root BH \\
\hline
0.30 & 0.00 & --- & No horizon (naked singularity) \\
\hline
0.30 & 0.50 & $1.30897$ & Extremal or Single Root BH \\
\hline
0.30 & 1.00 & $1.10520$ & Extremal or Single Root BH \\
\hline
0.70 & 0.00 & $0.285714$ & Extremal or Single Root BH \\
\hline
0.70 & 0.50 & $0.271636$ & Extremal or Single Root BH \\
\hline
0.70 & 1.00 & $0.234946$ & Extremal or Single Root BH \\
\hline
1.00 & 0.00 & $2.00000$ & Extremal or Single Root BH \\
\hline
1.00 & 0.50 & $0.133975,\ 1.86603$ & Non-extremal BH \\
\hline
1.00 & 1.00 & --- & No horizon (naked singularity) \\
\hline
1.20 & 0.00 & $2.66667$ & Extremal or Single Root BH \\
\hline
1.20 & 0.50 & $0.309216,\ 2.59881$ & Non-extremal BH \\
\hline
1.20 & 1.00 & $0.823399,\ 2.34440$ & Non-extremal BH \\
\hline
1.50 & 0.00 & $3.33333$ & Extremal or Single Root BH \\
\hline
1.50 & 0.50 & $0.435535,\ 3.30777$ & Non-extremal BH \\
\hline
1.50 & 1.00 & $0.841698,\ 3.22486$ & Non-extremal BH \\
\hline
2.00 & 0.00 & $4.00000$ & Extremal or Single Root BH \\
\hline
2.00 & 0.50 & $0.523956,\ 3.99214$ & Non-extremal BH \\
\hline
2.00 & 1.00 & $0.860440,\ 3.96799$ & Non-extremal BH \\
\hline
\caption{Event horizon configurations for various values of the non-minimal coupling parameter $k$ and electric charge $q$ in STPG spacetime. The classification is based on the nature and number of real, positive roots of the metric function $f(r)$ given in Eq.~\eqref{ismetric2}. The transition from naked singularities to single-horizon and dual-horizon configurations demonstrates the rich causal structure arising from the interplay between electromagnetic charge and non-minimal gravitational coupling in the STPG framework.}
\label{tab:horizons}
\end{longtable*}

\begin{figure*}
    \centering
\includegraphics[width=1\textwidth]{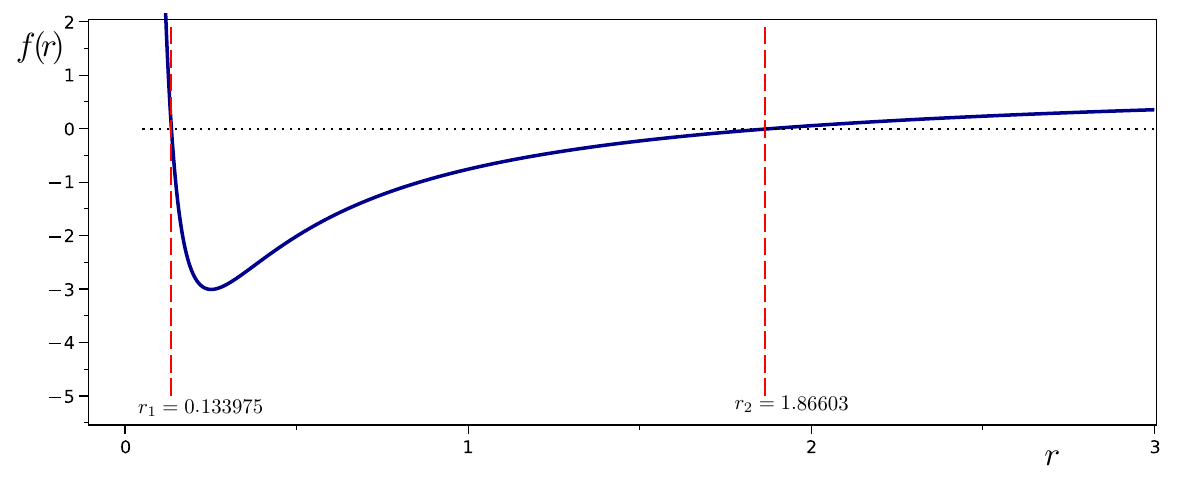}
    \caption{The metric function $f(r)$ plotted for the specific case $k=1$, $q=0.5$, and $M=1$ in the STPG BH spacetime. The two red dashed vertical lines indicate the horizon radii at $r_1 = 0.133975$ and $r_2 = 1.86603$, corresponding to a non-extremal BH configuration with inner (Cauchy) and outer (event) horizons respectively. The function exhibits the characteristic behavior of crossing zero at both horizons while maintaining positive values in the intermediate region, ensuring proper Lorentzian signature. The mass parameter is chosen as $M=1$ for illustration purposes, demonstrating how non-minimal electromagnetic coupling modifies the classical Schwarzschild structure.}
    \label{fig:metric_function_plot}
\end{figure*}

The behavior of the metric function $f(r)$ shown in Fig.~\ref{fig:metric_function_plot} illustrates the presence of two distinct horizons for the case $k=1$, $q=0.5$, and $M=1$. The function crosses the horizontal axis at $r_1 = 0.133975$ and $r_2 = 1.86603$, corresponding to the inner (Cauchy) and outer (event) horizons, respectively. This configuration confirms a non-extremal BH, where the spacetime is causally separated into distinct regions. The presence of two positive, real roots further supports the general conclusion that the non-minimal coupling can preserve standard BH structure for suitable ranges of $k$ and $q$. As $f(r)$ approaches zero at the horizons and remains positive in between, it ensures a Lorentzian signature outside the event horizon.

The consistency of the solution is verified through the constraint \cite{doyran2024non}:

\begin{equation}
\frac{d\mathcal{Y}}{dQ} E^2 = 1 - k,
\end{equation}

which is naturally satisfied when inserting the above expression of $f(r)$ into the definition of $Q$, explicitly:

\begin{equation}
Q = -\frac{1}{\kappa} \left( \frac{f{'}}{r} + \frac{f}{r^2} \right).
\end{equation}

Finally, we express the non-minimal coupling function $\mathcal{Y}(Q)$ through a compact integral relation:

\begin{equation}
\mathcal{Y}(Q) = 1 + (1 - k) \int \frac{dQ}{\mathcal{F}}.
\end{equation}

Inserting back into the original action gives the full Lagrangian:

\begin{eqnarray}
  \mathcal{L}  &=& Q*1 + \left[1 + (1 - k) \int \frac{dQ}{\mathcal{F}} \right] \mathcal{F}*1 + \Lambda \star 1 \nonumber \\
&&+ T^a \wedge \lambda_a + R^a_{\ b} \wedge \rho^b_{\ a}.  
\end{eqnarray}

This formulation emphasizes that the specific solution given by Eq.~\eqref{fr} is not an arbitrary metric ansatz but rather a direct consequence of a well-defined gravitational action with non-minimal coupling. It hence provides a bridge between metric-affine formulations and exact analytical solutions in modified gravity scenarios.

\subsection{Surface Gravity and Hawking Temperature of the STPG BH}

For the STPG spacetime described by the metric in Eq.~\eqref{fr}, the vector field $\xi = \frac{\partial}{\partial t}$ corresponds to a timelike Killing vector. Associated with this symmetry is a conserved quantity, which can be analyzed via the properties of $\xi^\mu$. The following identity relates the covariant derivative of the norm of the Killing vector to the surface gravity $\kappa$:

\begin{equation}
\nabla_\nu (\xi^\mu \xi_\mu) = -2 \kappa \xi_\nu.
\end{equation}

Here, $\nabla_\nu$ denotes the covariant derivative operator, and $\kappa$ remains constant along the flow lines of $\xi$, reflecting the stationarity of the spacetime. This is evident from the vanishing Lie derivative of $\kappa$ with respect to $\xi$:

\begin{equation}
\mathcal{L}_\xi \kappa = 0.
\end{equation}

The quantity $\kappa$, known as surface gravity, maintains a constant value across the BH horizon $r_h$. In the coordinate basis, the components of the Killing vector take the form $\xi^\mu = (1, 0, 0, 0)$. For the given BH metric, surface gravity is found by evaluating the radial derivative of the lapse function at the horizon:

\begin{equation}
\kappa = \frac{1}{2} f'(r) \Big|_{r = r_h} 
\end{equation}

Hawking's seminal discovery \cite{hawking1975particle} established that BHs radiate thermally with a temperature proportional to the surface gravity. Applying this relation, the Hawking temperature for the charged BH under consideration becomes:

\begin{equation}
T_{\text{H}} = \frac{\kappa}{2\pi} = \frac{M}{2 \pi r_h^{2}}-\frac{3 \zeta  q^{2} \left(r_h^{\frac{1}{k}}\right)^{4}}{2 \pi r_h^{7}}+\frac{\zeta  q^{2} \left(r_h^{\frac{1}{k}}\right)^{4}}{\pi r_h^{7} k}.\label{hawkingg}
\end{equation}

Equation \eqref{hawkingg} reveals several important features of the STPG BH thermal behavior.  The first term represents the classical gravitational contribution proportional to the mass $M$, while the second and third terms arise from the non-minimal electromagnetic coupling and depend critically on the coupling parameter $k$. The fractional power $(r_h^{1/k})^4$ in the charge-dependent terms demonstrates how the non-minimal coupling modifies the standard Coulomb-like behavior, creating a more complex radial dependence than in conventional Reissner-Nordstr\"{o}m BHs. The competing signs of the electromagnetic terms suggest the possibility of temperature extrema as functions of the horizon radius, indicating a special thermodynamic behavior unique to the STPG framework.

\begin{figure}
    \centering
    \includegraphics[width=0.55\textwidth]{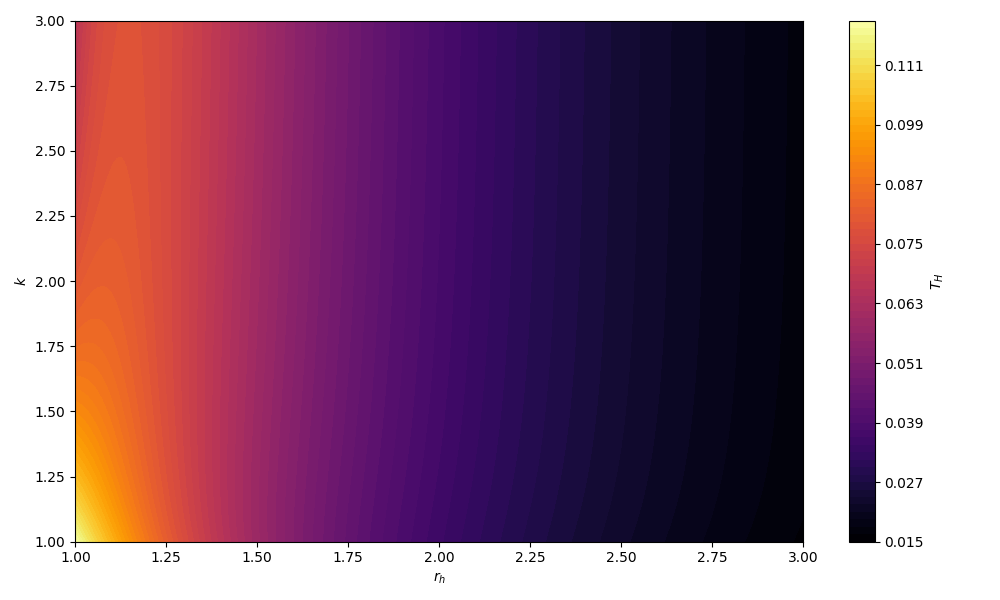}
    \caption{Variation of the Hawking temperature $T_H$ with respect to the non-minimal coupling parameter $k$ for the STPG BH spacetime for fixed parameters are $M=1$, $q=0.5$, $\zeta=1$, demonstrating how electromagnetic-gravitational coupling strength affects the thermal properties of the BH.}
    \label{hawkingfig}
\end{figure}

\begin{figure}
    \centering
    \includegraphics[width=0.55\textwidth]{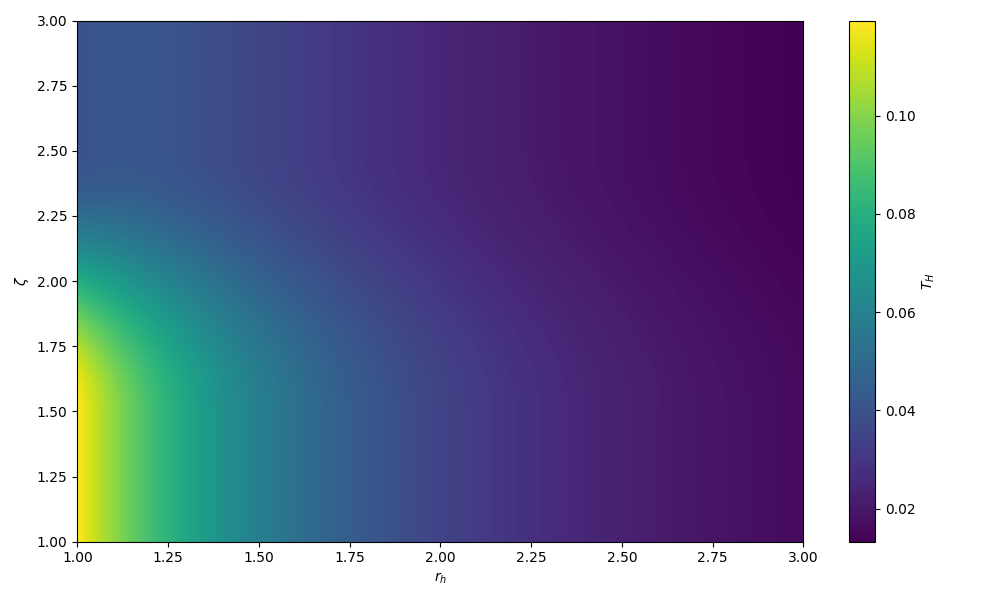}
    \caption{Variation of the Hawking temperature $T_H$ with respect to the gravitational coupling constant $\zeta$ for the specific case for $k=1$, $M=1$, and $q=0.5$  in STPG framework. Illustrating the dependence of BH thermal radiation on the strength of gravitational interactions in the STPG framework.}
    \label{hawkingfig2}
\end{figure}

\section{Quantum-Corrected Thermodynamics of the   BH\lowercase{s}} \label{sec3}

As BHs evaporate through Hawking radiation, their thermodynamic structure demands generalization beyond the classical framework, especially when their characteristic sizes approach the Planck length order. In this quantum regime, microstate configurations and fluctuations inaccessible in the semiclassical regime begin to matter significantly. Among the numerous quantities affected is the entropy, classically proportional to the surface area of the event horizon by the Bekenstein-Hawking formula.

To incorporate quantum features, statistical mechanics can be utilized through counting microscopic configurations pertaining to the BH horizon \cite{ghosh2006counting,krasnov1998quantum}. In these approaches, the entropy is derived from the logarithm of microstates available in equilibrium. When considering quantum fluctuations in this counting, especially under total fixed-energy and particle-number conditions, the statistical entropy acquires subleading corrections. These corrections reflect the probabilistic nature of the underlying quantum system and encode deviations from the leading area law.

Among various candidates for such corrections, an exponential form has been discovered in some detailed microstate models, where the entropy is corrected by a term that possesses an exponential decay in the classical entropy \cite{chatterjee2020exponential}. This leads to the corrected entropy expression. Among various possible correction forms (e.g., logarithmic, power-law), we select the exponential term due to its emergence in certain microstate counting approaches \cite{gursel2025thermodynamics}, and because it decays more rapidly for large $S_0$, ensuring negligible influence in the macroscopic BH regime while dominating the microscopic limit.\cite{chatterjee2020exponential, ali2024quantum,sucu2025exploring}:

\begin{equation}
S = S_0 + e^{-S_0}. \label{expcorrected}
\end{equation}

Here, $S_0$ is the uncorrected Bekenstein-Hawking entropy and the additive term $e^{-S_0}$ captures leading-order quantum contributions and becomes significant in the regime of small BH radii. The exponential correction form $e^{-S_0}$ is chosen over alternative quantum corrections such as logarithmic terms $\ln(S_0)$ or power-law modifications for several theoretical and practical reasons \cite{isrply29,isrply30}. In microstate counting approaches to BH entropy, the exponential correction emerges naturally from statistical mechanical considerations when accounting for finite-size effects and correlation functions between horizon degrees of freedom \cite{isrply31,isrply32}. Unlike logarithmic corrections that grow without bound, the exponential term $e^{-S_0}$ exhibits rapid decay for large classical entropy $S_0$, ensuring that quantum effects remain negligible in the macroscopic BH regime while dominating for microscopic BHs approaching Planck scales. This asymptotic behavior is physically motivated since quantum corrections should vanish in the classical limit $S_0 \gg 1$, while becoming significant when $S_0 \sim 1$. The exponential form also preserves the thermodynamic stability conditions and ensures finite corrections to all derived quantities, unlike certain logarithmic corrections that can introduce divergences in thermodynamic derivatives. Additionally, exponential corrections have been identified in loop quantum gravity approaches and in certain string theory models for BH microstates \cite{isrply33,isrply34,isrply35}. The specific form $S = S_0 + e^{-S_0}$ represents the leading-order quantum correction in this framework, with higher-order terms being suppressed by additional powers of $e^{-S_0}$.

To derive the quantum-corrected internal energy, we employ the fundamental thermodynamic relation $dE = TdS - PdV + \Phi dq$, where $\Phi$ is the electric potential. For our BH system with vanishing pressure ($P = 0$), this reduces to \cite{pourhassan2022exponential}:

\begin{equation}
dE_C = T_H dS = T_H \left( dS_0 + d(e^{-S_0}) \right).
\end{equation}

Using the chain rule for the exponential correction term:

\begin{equation}
d(e^{-S_0}) = -e^{-S_0} dS_0,
\end{equation}

the corrected energy differential becomes:

\begin{equation}
dE_C = T_H (1 - e^{-S_0}) dS_0
\end{equation}

Integrating this expression with $S_0 = \pi r_h^2$ and the Hawking temperature from Eq.~\eqref{hawkingg} yields the internal energy as follows
\begin{multline}
E_{C} \approx \frac{\pi  M \,r_h^{2}}{2}+\frac{3 \pi  \zeta  q^{2} \left(r_h^{\frac{1}{k}}\right)^{4} k}{\left(-4+3 k\right) r_h^{3}}-\frac{2 \pi  \zeta  q^{2} \left(r_h^{\frac{1}{k}}\right)^{4}}{\left(-4+3 k\right) r_h^{3}}. \label{s42}
\end{multline}

\begin{figure}
    \centering
    \includegraphics[width=0.5\textwidth]{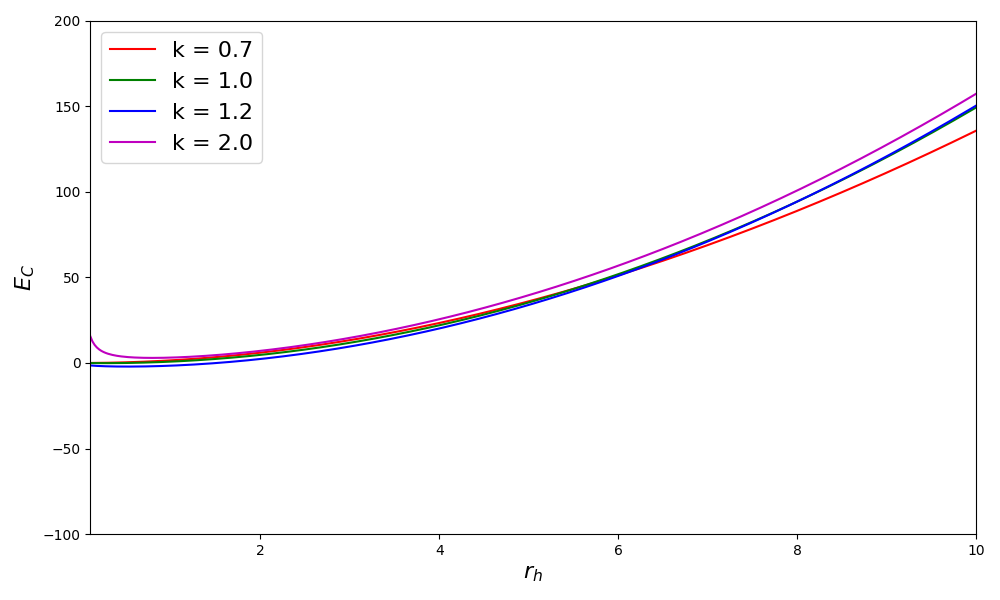}
    \caption{Quantum-corrected internal energy $E_C$ versus horizon radius $r_h$  for $k \in \{0.7, 1.0, 1.2, 1.5, 2.0\}$, with fixed $q=0.5$, $\zeta=1$, and $M=1$.}
    \label{enerji}
\end{figure}

The behavior shown in Fig.~\ref{enerji} reveals the variation of the corrected energy $E_C$ with the event horizon radius $r_h$ for different values of $k$. For small values of $r_h$, the energy approaches a minimum, and for some values of $k$, a negative energy regime is observed, indicating that quantum corrections or coupling effects dominate for small BHs. In the large $r_h$ regime, all curves increase in a positive direction, indicating a return to classical Schwarzschild-like behavior. Furthermore, the upward trend of the energy curve as $k$ increases indicates that stronger coupling increases the energy of the system and that the effect of the electromagnetic field on the thermodynamic structure becomes more pronounced \cite{javed2023thermal}. These observations demonstrate that the phase structure of the system is sensitive to $k$ and that critical behaviors can be controlled by parameter settings.

Similarly, the corrected Helmholtz free energy is obtained via:

\begin{equation}
F_{C} = -\int S dT_H, \label{s43}
\end{equation}

which, using Eqs.~\eqref{expcorrected} and \eqref{hawkingg}, yields:

\begin{multline}
F_{C} \approx \frac{21 \pi \zeta  q^{2} \left(r_h^{\frac{1}{k}}\right)^{4} k}{4 \left(-4+3 k\right) r_h^{3}}-\frac{13 \pi  \zeta  q^{2} \left(r_h^{\frac{1}{k}}\right)^{4}}{2 \left(-4+3 k\right) r_h^{3}}\\+\frac{9 k \,q^{2} \zeta  \left(r_h^{\frac{1}{k}}\right)^{4}}{2 r_h^{7} \pi  \left(-4+3 k\right)}-\frac{9 q^{2} \zeta  \left(r_h^{\frac{1}{k}}\right)^{4}}{r_h^{7} \pi  \left(-4+3 k\right)}\\+\frac{2 \pi  q^{2} \zeta  \left(r_h^{\frac{1}{k}}\right)^{4}}{r_h^{3} k \left(-4+3 k\right)}+\frac{4 q^{2} \zeta  \left(r_h^{\frac{1}{k}}\right)^{4}}{r_h^{7} k \pi  \left(-4+3 k\right)}\\-\frac{r_h^{2} \pi  M}{-4+3 k}+\frac{2 M}{r_h^{2} \pi  \left(-4+3 k\right)}\\+\frac{3 r_h^{2} k \pi  M}{4 \left(-4+3 k\right)}-\frac{3 k M}{2 r_h^{2} \pi  \left(-4+3 k\right)}\label{free}
\end{multline}

\begin{figure}
    \centering
    \includegraphics[width=0.5\textwidth]{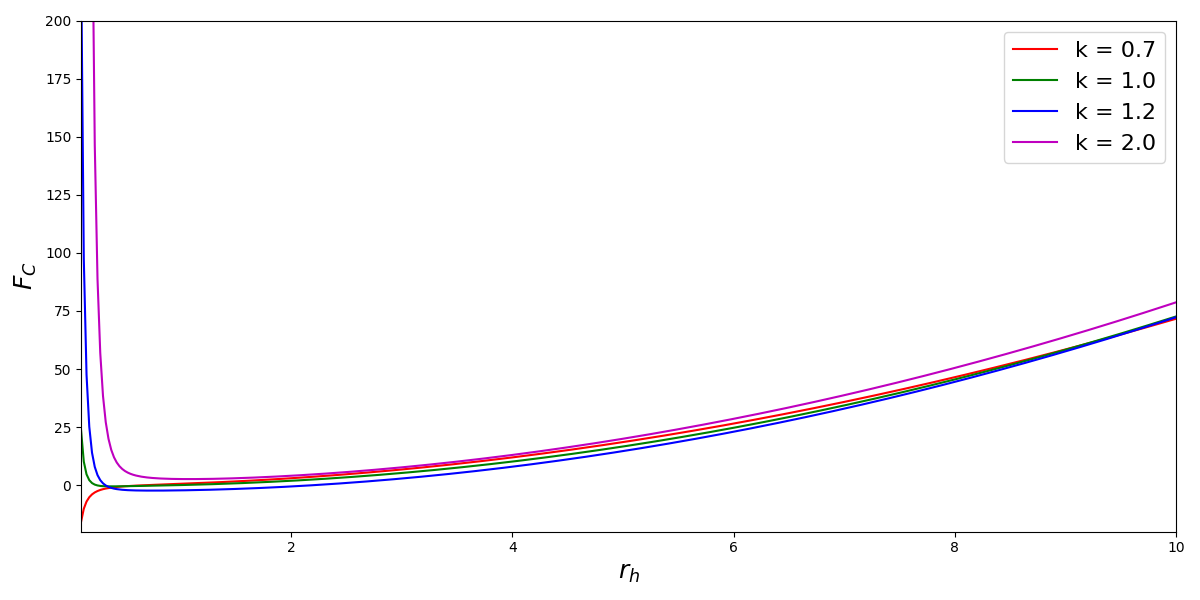}
    \caption{Behavior of the quantum-corrected Helmholtz free energy $F_C$ as a function of the horizon radius $r_h$ for STPG BHs, plotted for different non-minimal coupling values $k$  under fixed parameters $q = 0.5$, $\zeta = 1$, and $M = 1$. The steep rise in $F_C$ at small horizon sizes for higher $k$ illustrates the thermodynamic cost induced by strong gravity-electromagnetic coupling in the quantum BH regime, demonstrating how non-minimal coupling affects the stability of microscopic BH configurations.}
    \label{freeF}
\end{figure}

The behavior depicted in Fig.~\ref{freeF} shows the variation of the corrected Helmholtz free energy $F_C$ with the event horizon radius $r_h$ for various values of $k$. In the small $r_h$ regime, especially for $k > 1$, the $F_C$ curves reach very high values and diverge in a positive direction, indicating that the system is unstable in this region and that significant deviations from classical BH thermodynamics are occurring. This behavior suggests that small BHs can be thermodynamically suppressed by quantum corrections or non-minimal coupling effects. As $r_h$ increases, all curves become positive and convex, while the divergence of the curves for different values of $k$ reflects the sensitivity of the free energy to the coupling strength. At larger values of $k$, $F_C$ is at higher levels, demonstrating that the electromagnetic-gravitational interaction significantly increases the Helmholtz potential and reduces the thermodynamic freedom of the system. These results emphasize that the equilibrium and phase structure of BHs are sensitively determined not only by the geometry but also by the level of interaction with the electromagnetic sector.

The pressure under quantum corrections is then computed as:

\begin{equation}
P_{C} = -\frac{d F_{C}}{d V}, \label{pressure}
\end{equation}

leading to the expression:

\begin{multline}
P_{C} \approx \frac{21 q^{2} \zeta  \left(r_h^{\frac{1}{k}}\right)^{4}}{16 r_h^{6}}-\frac{13 q^{2} \zeta  \left(r_h^{\frac{1}{k}}\right)^{4}}{8 r_h^{6} k}+\frac{q^{2} \zeta  \left(r_h^{\frac{1}{k}}\right)^{4}}{2 r_h^{6} k^{2}}\\+\frac{21 q^{2} \zeta  \left(r_h^{\frac{1}{k}}\right)^{4}}{8 \pi^{2} r_h^{10}}-\frac{13 q^{2} \zeta  \left(r_h^{\frac{1}{k}}\right)^{4}}{4 \pi^{2} r_h^{10} k}\\+\frac{q^{2} \zeta  \left(r_h^{\frac{1}{k}}\right)^{4}}{\pi^{2} r_h^{10} k^{2}}-\frac{M}{8 r_h}-\frac{M}{4 \pi^{2} r_h^{5}}\label{pres}
\end{multline}

\begin{figure}
    \centering
    \includegraphics[width=0.5\textwidth]{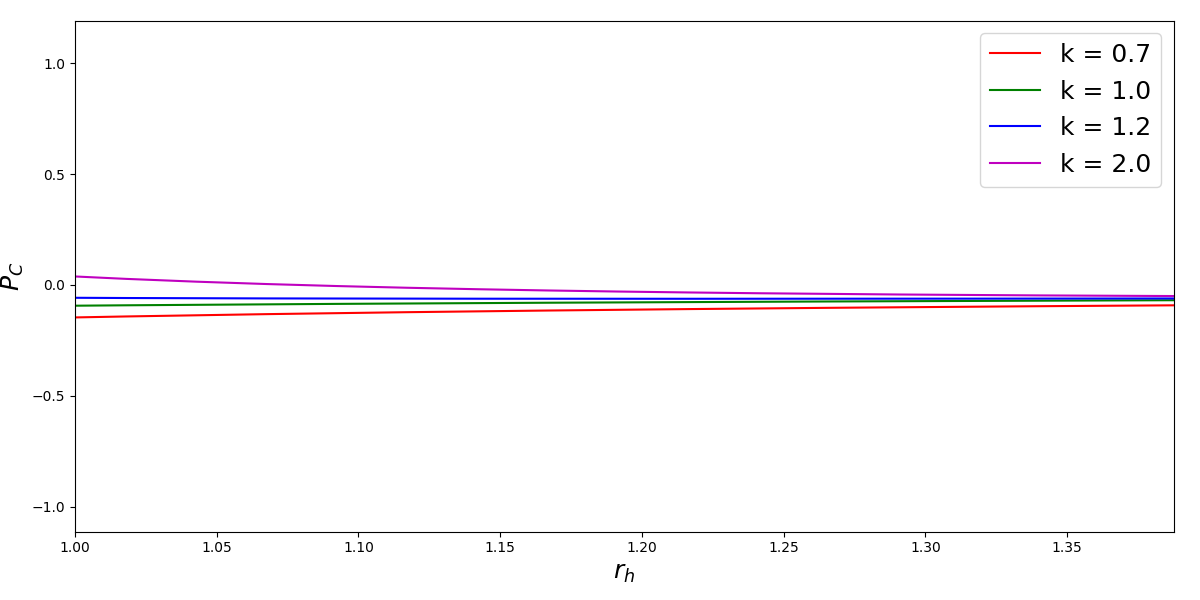}
    \caption{Quantum-corrected pressure $P_C$ as a function of horizon radius $r_h$ for STPG BHs with various coupling parameters $k$, demonstrating the complex pressure behavior arising from the interplay between quantum corrections and non-minimal electromagnetic coupling effects, with fixed charge $q=0.5$, \textcolor{red}{M=1 and $\zeta$=1.}}
    \label{pressureP}
\end{figure}

By combining various thermodynamic variables, we obtain the enthalpy $H_C$, which is given by:

\begin{equation}
H_C=E_C+P_CV. \label{s35}
\end{equation}

Equation \eqref{s35} reduces to the following when the associated variables are substituted:

\begin{multline}
H_C \approx -\frac{27 q^{2} \zeta  \left(r_h^{\frac{1}{k}}\right)^{4}}{r_h^{7} \pi \left(-4+3 k\right)}+\frac{64 q^{2} \zeta  \left(r_h^{\frac{1}{k}}\right)^{4}}{3 r_h^{7} k \pi \left(-4+3 k\right)}\\-\frac{16 q^{2} \zeta  \left(r_h^{\frac{1}{k}}\right)^{4}}{3 \pi r_h^{7} k^{2} \left(-4+3 k\right)}+\frac{21 k \,q^{2} \zeta  \left(r_h^{\frac{1}{k}}\right)^{4}}{2 r_h^{7} \pi \left(-4+3 k\right)}\\-\frac{31 \pi \zeta  q^{2} \left(r_h^{\frac{1}{k}}\right)^{4}}{2 \left(-4+3 k\right) r_h^{3}}+\frac{32 \pi q^{2} \zeta  \left(r_h^{\frac{1}{k}}\right)^{4}}{3 r_h^{3} k \left(-4+3 k\right)}\\-\frac{8 \pi q^{2} \zeta  \left(r_h^{\frac{1}{k}}\right)^{4}}{3 r_h^{3} k^{2} \left(-4+3 k\right)}+\frac{33 \pi \zeta  q^{2} \left(r_h^{\frac{1}{k}}\right)^{4} k}{4 \left(-4+3 k\right) r_h^{3}}\\-\frac{4 r_h^{2} \pi M}{3 \left(-4+3 k\right)}+\frac{4 M}{3 r_h^{2} \pi \left(-4+3 k\right)}\\+\frac{r_h^{2} k \pi M}{-4+3 k}-\frac{k M}{r_h^{2} \pi \left(-4+3 k\right)}. \label{s36}
\end{multline}

\begin{figure}
    \centering
    \includegraphics[width=0.5\textwidth]{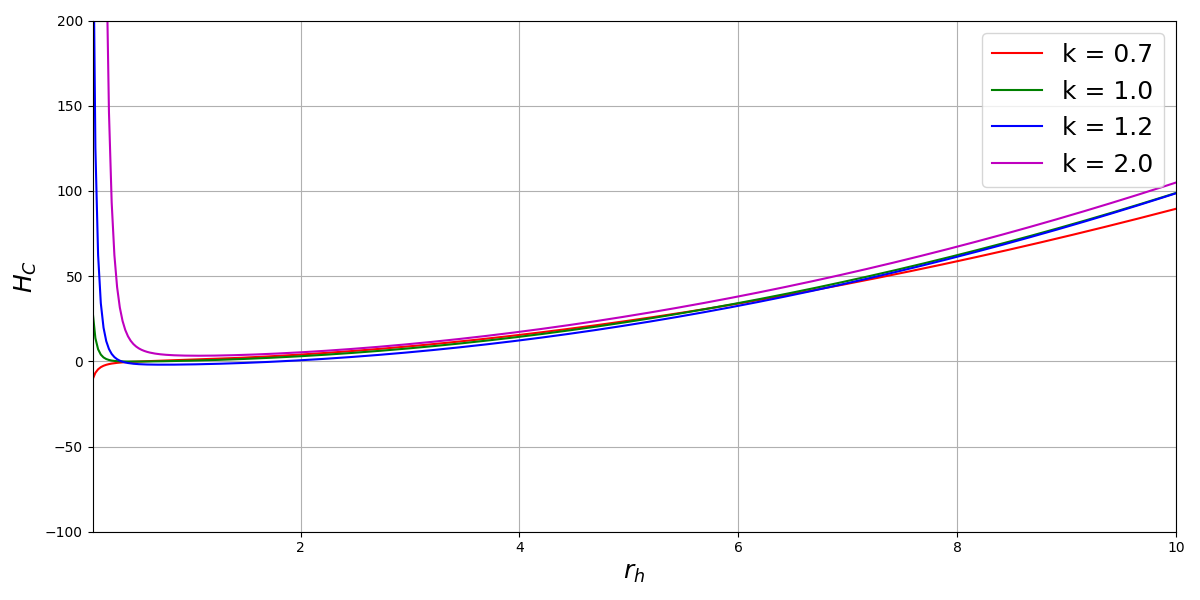}
    \caption{Quantum-corrected enthalpy $H_C$  versus horizon radius $r_h$  for different coupling parameters $k$, with fixed $q=0.5$, $\zeta=1$, and $M=1$.}
    \label{gibbs_H}
\end{figure}

In Fig.~\ref{gibbs_H}, the relationship between the BH's enthalpy $H_C$ and the event horizon radius $r_h$ is examined for different coupling parameters $k$. The graph generally shows that $H_C$ increases with $r_h$, meaning that as the BH's size increases, its internal energy and hence its thermodynamic potential enthalpy also increases. However, in the small $r_h$ regime, a distinct peak in enthalpy (divergent behavior) is observed, especially for $k > 1$; this suggests that in highly coupled states, small-horizon BHs may be thermodynamically unstable or susceptible to quantum effects. This behavior suggests that in strong coupling regimes, the BH's internal energy can change rapidly due to interaction with the electromagnetic field, and classical thermodynamic assumptions may break down there. As a result, increasing the coupling parameter $k$ sharpens the increase in enthalpy rate, especially in small regions $r_h$, providing important insights into the microscopic nature of the system.

The Gibbs free energy is one significant state function that is necessary to ascertain the BH's global stability. Using $G = H - TS$, we can find the relation for Gibbs free energy, and we obtain:

\begin{multline}
    G_{C} \approx \frac{21 \pi \zeta  q^{2} \left(r_h^{\frac{1}{k}}\right)^{4} k}{2 \left(-4+3 k\right) r_h^{3}}-\frac{20 \pi \zeta  q^{2} \left(r_h^{\frac{1}{k}}\right)^{4}}{\left(-4+3 k\right) r_h^{3}}\\+\frac{38 \pi q^{2} \zeta  \left(r_h^{\frac{1}{k}}\right)^{4}}{3 r_h^{3} k \left(-4+3 k\right)}-\frac{8 \pi q^{2} \zeta  \left(r_h^{\frac{1}{k}}\right)^{4}}{3 r_h^{3} k^{2} \left(-4+3 k\right)}\\-\frac{36 q^{2} \zeta  \left(r_h^{\frac{1}{k}}\right)^{4}}{r_h^{7} \pi \left(-4+3 k\right)}+\frac{76 q^{2} \zeta  \left(r_h^{\frac{1}{k}}\right)^{4}}{3 r_h^{7} k \pi \left(-4+3 k\right)}\\-\frac{16 q^{2} \zeta  \left(r_h^{\frac{1}{k}}\right)^{4}}{3 \pi r_h^{7} k^{2} \left(-4+3 k\right)}-\frac{r_h^{2} \pi M}{3 \left(-4+3 k\right)}\\+\frac{10 M}{3 r_h^{2} \pi \left(-4+3 k\right)}+\frac{15 k \,q^{2} \zeta  \left(r_h^{\frac{1}{k}}\right)^{4}}{r_h^{7} \pi \left(-4+3 k\right)}\\+\frac{r_h^{2} k \pi M}{4 \left(-4+3 k\right)}-\frac{5 k M}{2 r_h^{2} \pi \left(-4+3 k\right)}
\end{multline}

Lastly, the quantum-corrected heat capacity is given by \cite{sucu2025quantum}:

\begin{equation}
C_{C} = T_H \left( \frac{\partial S}{\partial T_H} \right), \label{s25}
\end{equation}

which, upon using the entropy and temperature relations, evaluates to:

\begin{equation}
C_{C} \approx -\frac{2 \pi^{2} \left(-3 q^{2} \left(k-\frac{2}{3}\right) \zeta  r_h^{\frac{-5 k+4}{k}}+M k\right) r_h^{9} k}{-21 q^{2} \left(k-\frac{4}{7}\right) \left(k-\frac{2}{3}\right) \zeta  r_h^{\frac{4}{k}}+2 M \,r_h^{5} k^{2}} . \label{heaaat}
\end{equation}

\begin{figure}
    \centering
    \includegraphics[width=0.5\textwidth]{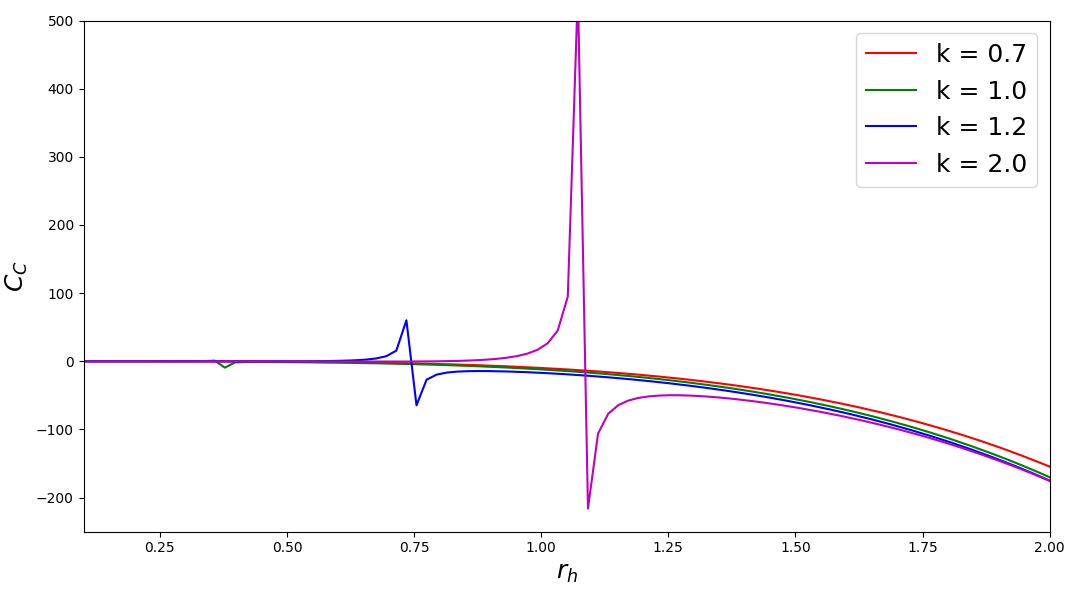}
    \caption{Quantum-corrected heat capacity $C_C$ plotted against the event horizon radius $r_h$ for STPG BHs with various coupling strengths $k$, assuming $q = 0.5$, $\zeta = 1$, and $M = 1$. The divergence points on the curves for $k = 1.2$ and $k = 2.0$ signal second-order phase transitions, indicating the existence of critical radii beyond which the thermodynamic stability of the BH drastically changes. The sign changes in $C_C$ reveal regions of thermodynamic stability (positive) versus instability (negative), demonstrating the rich phase structure of quantum-corrected STPG BHs.}
    \label{heattttt}
\end{figure}

Figure~\ref{heattttt} shows the variation of the corrected heat capacity $C_C$ with the event horizon radius $r_h$ for different coupling strengths ($k$). The graph exhibits striking phase transitions, especially in the small $r_h$ range. Vertical asymptotes (divergences) appear in $C_C$ as $r_h$ approaches certain critical values for $k = 1.2$ and $k = 2.0$, indicating second-order phase transitions in the system. These points correspond to the critical horizon radii where abrupt changes in the thermodynamic stability of the BH occur. Positive $C_C$ corresponds to regions where the system is thermodynamically stable, while negative $C_C$ corresponds to regimes of instability. This critical behavior is particularly pronounced for large $k$ values, occurring at higher $r_h$ positions, clearly demonstrating the influence of electromagnetic-gravitational coupling on the phase structure. In summary, the critical behavior of the system can be precisely controlled by $k$, revealing the decisive role of non-minimal coupling on phase transitions in quantum-corrected STPG BHs.

\section{Quantum-Corrected JTE Analysis for  STPG BHs} \label{sec4}

The JTE is a thermodynamic process where the enthalpy remains constant while temperature varies with pressure. In the context of BHs, this effect provides an isenthalpic pathway within the extended phase space \cite{yekta2019joule,bi2021joule,cao2022black}. The Joule-Thomson coefficient, which determines the temperature variation with pressure, indicates whether the system undergoes heating or cooling during expansion. This analysis becomes particularly intriguing when quantum corrections are incorporated into the STPG framework \cite{wang2025thermodynamic,chaudhary2022thermodynamic}.

The heat capacity of the BH, derived from the first law of thermodynamics, is expressed as Eq.~\eqref{s25} using the entropy and temperature relations. The specific heat obtained in Eq.~\eqref{heaaat} from earlier sections provides the foundation for our JTE analysis in the quantum-corrected regime.

The Joule-Thomson coefficient, which measures how temperature responds to pressure during isenthalpic expansion, is defined as \cite{mo2018joule}:

\begin{equation}
\mu_J = \left( \frac{\partial T_H}{\partial P_C} \right)_H = \frac{\left( \frac{\partial T_H}{\partial r_h} \right)}{\left( \frac{\partial P_C}{\partial r_h} \right)}, \label{isadd1}
\end{equation}

At the horizon, the condition $f(r_h) = 0$ from Eq.~\eqref{ismetric2} gives us:

\begin{equation}
\frac{k}{3k-2} - \frac{2M}{r_h} + \frac{q^2}{r_h^{\frac{6k-4}{k}}} = 0. \label{isadd2}
\end{equation}

Solving for $M$ in terms of $r_h$:

\begin{equation}
M = \frac{r_h}{2} \left( \frac{k}{3k-2} + \frac{q^2}{r_h^{\frac{6k-4}{k}}} \right). \label{isadd3}
\end{equation}

Substituting this relation into the Hawking temperature expression and computing the derivative, one finds out

\begin{equation}
\frac{\partial T_H}{\partial r_h} = \frac{\partial}{\partial r_h} \left[ \frac{1}{2\pi} \left( \frac{M}{r_h^2} - \frac{3\zeta q^2 r_h^{\frac{4}{k}}}{2r_h^7} + \frac{\zeta q^2 r_h^{\frac{4}{k}}}{kr_h^7} \right) \right]. \label{isadd3}
\end{equation}

After algebraic manipulation and using the horizon condition, this yields the numerator:

\begin{equation}
X = 4k \left( -21 \left(k-\frac{2}{3}\right) \left(k-\frac{4}{7}\right) \zeta q^2 r_h^{\frac{-5k+4}{k}} + 2M k^2 \right) r_h^8 \pi. \label{isadd4}
\end{equation}

Similarly, the pressure derivative calculation involves:

\begin{equation}
\frac{\partial P_C}{\partial r_h} = -\frac{\partial}{\partial r_h} \left[ \frac{\partial F_C}{\partial V} \right], \label{isadd5}
\end{equation}

where the volume is related to the horizon radius as $V = \frac{4\pi r_h^3}{3}$. The detailed calculation of mixed partial derivatives, combined with the horizon condition and quantum corrections, leads to the denominator:

\begin{multline}
Y = -63 \left(k-\frac{2}{3}\right)^2 \left(k-\frac{4}{7}\right) \zeta q^2 \pi^2 r_h^{\frac{4k+4}{k}} \\
-210 \left(k-\frac{2}{3}\right) \left(k-\frac{4}{7}\right) \zeta \left(k-\frac{2}{5}\right) q^2 r_h^{\frac{4}{k}} \\
+ M r_h^5 k^3 \left( \pi^2 r_h^4 + 10 \right). \label{isadd6}
\end{multline}

The Joule-Thomson coefficient is then expressed as $\mu_J = -X/Y$, where the specific fractional terms $(k-2/3)$, $(k-4/7)$, and $(k-2/5)$ arise from the interplay between quantum corrections and the non-minimal electromagnetic coupling in the STPG framework.  

\begin{figure}
    \centering
    \includegraphics[width=0.5\textwidth]{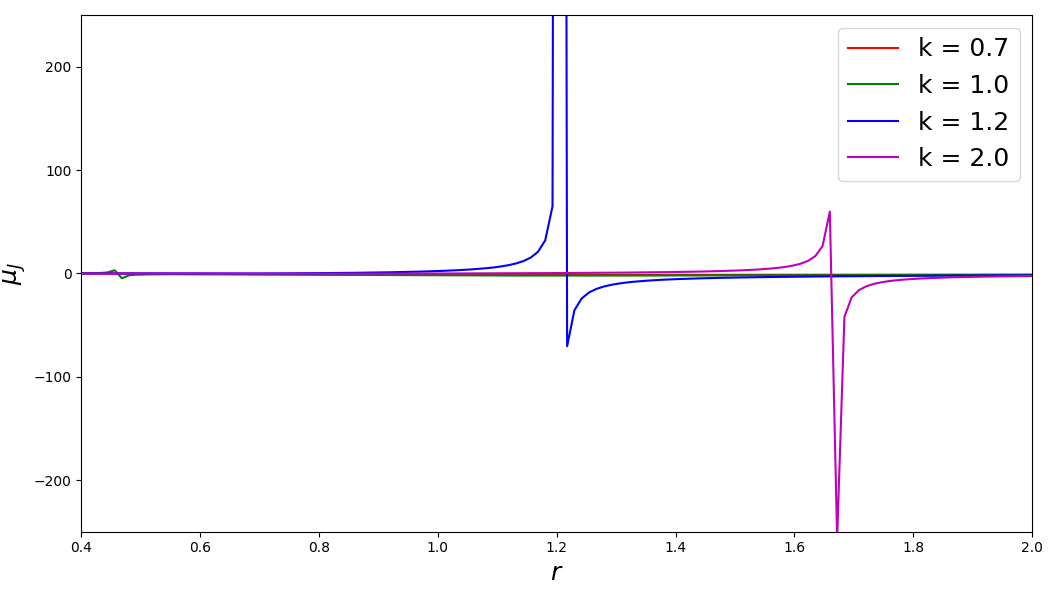}
    \caption{Quantum-corrected Joule-Thomson coefficient $\mu_J$ versus event horizon radius $r_h$ for STPG BHs with different non-minimal coupling parameters $k$ and fixed charge $q=0.5$ $\zeta=1$, and $M=1$. The plot reveals the effects of non-minimal electromagnetic coupling on the thermodynamic cooling-heating behavior of the BH. Each curve exhibits a divergence at a critical radius $r_h = r_{ci}$ (inversion point), where $\mu_J$ transitions from positive (cooling regime) to negative (heating regime). The shift of inversion points to smaller $r_h$ values for larger $k$ indicates enhanced thermodynamic instability under strong coupling effects in the quantum-corrected regime.}
    \label{JTE}
\end{figure}

The behavior shown in Fig.~\ref{JTE} demonstrates the Joule-Thomson coefficient $\mu_J$ for different non-minimal coupling parameters $k$ as a function of the event horizon radius $r_h$, clearly revealing the effects of non-minimal electromagnetic coupling on the thermodynamic cooling-heating behavior of STPG BHs. The sign of $\mu_J$ indicates the thermodynamic phase: cooling for positive values and heating for negative ones \cite{javed2020effect}, with this behavior being particularly rich in the quantum-corrected STPG framework. Each curve exhibits a polar divergence at a critical radius $r_h = r_{ci}$, where $\mu_J$ transitions from positive to negative for a given value of $k$, representing the inversion point of the JTE.

In the regions $r_h < r_{ci}$, $\mu_J > 0$ indicates that the BH will cool down during isenthalpic expansion, while in the regions $r_h > r_{ci}$, $\mu_J < 0$ indicates that it will heat up. At larger values of $k$, the divergence points shift to smaller values of $r_h$ and the curves show a sharper transition, indicating that the BH becomes more prone to thermodynamic instability under the strong non-minimal effects of the electromagnetic field. The cooling region narrows, and the JTE occurs in a more limited range of effects. This behavior is consistent with the physical properties expected at high energy densities and demonstrates the unique signatures of quantum-corrected STPG BHs \cite{wang2025thermodynamic,gracca2023joule}.

Setting $\mu_J = 0$ gives the inversion point, from which the inversion temperature $T_i$ can be calculated using:

\begin{equation}
T_i = V \left( \frac{\partial T}{\partial V} \right)_P.
\end{equation}

Solving $\mu_J = 0$ analytically leads to the inversion horizon radius $r_{ci}$ through the condition that determines the transition between cooling and heating regimes in the quantum-corrected STPG framework. This inversion point provides crucial information about the thermodynamic stability and phase structure of these modified gravity BHs, offering potential observational signatures that could distinguish STPG from GR predictions.

\section{Thermodynamic Efficiency of STPG BH Heat Engines} \label{sec5}

In this section, we analyze the thermodynamic efficiency of a BH heat engine operating within the STPG framework in the case where the cosmological constant vanishes, i.e., the AdS background is absent \cite{xu2017black}. Consequently, the pressure $P$ in the extended phase space is zero, and the Hawking temperature is reduced to a simplified form that depends solely on the radius of the horizon $r_+$ and the electric charge $q$. This configuration provides a unique opportunity to study heat engine performance in modified gravity theories without the complications of AdS thermodynamics \cite{kruglov2024heat}.

Recall that the Hawking temperature in this context is derived from the surface gravity at the event horizon, as described by Eq. \eqref{hawkingg}. To assess the engine's performance, we consider a classical Carnot cycle where the BH undergoes a sequence of quasi-static processes between two temperatures, $T_H$ (hot reservoir) and $T_C$ (cold reservoir) \cite{fernando2018massive}. The Carnot cycle represents the most efficient heat engine operating between two thermal reservoirs and provides an upper bound for the efficiency of any real heat engine.

The Carnot efficiency is defined as \cite{fernando2018massive}:

\begin{equation}
\eta_C = 1 - \frac{T_C}{T_H}
\end{equation}

This result illustrates the fundamental limitation imposed by the temperature ratio of the BH's isothermal boundaries. Moreover, it underscores how the interplay between electric charge and non-minimal electromagnetic coupling governs the thermodynamic response of the system, even in the absence of a confining AdS potential \cite{bahamonde2025coupling,jana2024non}. The analysis confirms that the STPG BH can act as a physically viable thermodynamic engine with a finite and well-behaved efficiency.

\begin{table}
\centering

\begin{tabular}{|c|c|c|c|c|c|c|}
\hline
$k$ & $q$ & $r_h$ & $r_C$ & $T_H$ & $T_C$ & $\eta_C = 1 - \frac{T_C}{T_H}$ \\
\hline
0.7 & 0.0 & 1 & 2  & 0.15915 & 0.03979 & 0.75000 \\
0.7 & 0.0 & 1 & 5  & 0.15915 & 0.00637 & 0.96000 \\
0.7 & 0.0 & 1 & 10 & 0.15915 & 0.00159 & 0.99000 \\
0.7 & 0.0 & 2 & 5  & 0.03979 & 0.00637 & 0.84000 \\
0.7 & 0.0 & 2 & 10 & 0.03979 & 0.00159 & 0.96000 \\
0.7 & 0.0 & 5 & 10 & 0.00637 & 0.00159 & 0.75000 \\
\hline
0.7 & 0.5 & 1 & 2  & 0.15347 & 0.03746 & 0.75593 \\
0.7 & 0.5 & 1 & 5  & 0.15347 & 0.00565 & 0.96320 \\
0.7 & 0.5 & 1 & 10 & 0.15347 & 0.00130 & 0.99155 \\
0.7 & 0.5 & 2 & 5  & 0.03746 & 0.00565 & 0.84920 \\
0.7 & 0.5 & 2 & 10 & 0.03746 & 0.00130 & 0.96537 \\
0.7 & 0.5 & 5 & 10 & 0.00565 & 0.00130 & 0.77035 \\
\hline
0.7 & 1.0 & 1 & 2  & 0.13642 & 0.03046 & 0.77669 \\
0.7 & 1.0 & 1 & 5  & 0.13642 & 0.00350 & 0.97438 \\
0.7 & 1.0 & 1 & 10 & 0.13642 & 0.00041 & 0.99697 \\
0.7 & 1.0 & 2 & 5  & 0.03046 & 0.00350 & 0.88527 \\
0.7 & 1.0 & 2 & 10 & 0.03046 & 0.00041 & 0.98641 \\
0.7 & 1.0 & 5 & 10 & 0.00350 & 0.00041 & 0.88157 \\
\hline
\end{tabular}
\caption{Carnot efficiency values for different non-minimal coupling parameter $k$, electric charge $q$, hot reservoir horizon radius $r_h$, and cold reservoir horizon radius $r_C$ based on the Hawking temperature $T(r_h)$ derived from the STPG metric function. The efficiency calculations demonstrate how electromagnetic charge and non-minimal coupling affect the thermodynamic performance of STPG BH heat engines, with higher charges and larger temperature differences yielding superior efficiency values approaching the theoretical maximum.}
\label{tab:carnot-efficiency}
\end{table}

Table~\ref{tab:carnot-efficiency} presents the Carnot efficiency based on the Hawking temperature $\eta_C = 1 - \frac{T_C}{T_H}$ for different event horizon ($r_H$) and cold reservoir horizon ($r_C$) radii, electric charge ($q$), and non-minimal coupling parameter ($k = 0.7$). The table reveals several important features of STPG BH heat engines. The Carnot efficiency increases significantly with both $r_C$ and $q$. In particular, for fixed $r_h = 1$, as $r_C$ becomes larger (i.e., as the temperature difference between the hot and cold reservoirs increases), $\eta_C$ approaches values between 75\% and 99\%. This confirms that, as in the classical Carnot cycle, the thermal efficiency increases as the temperature difference between the hot and cold reservoirs increases.

Furthermore, as the electric charge increases (from $q = 0$ to $0.5$ to $1.0$), for the same values of $r_H$ and $r_C$, $T_C$ decreases further, thus increasing the efficiency. This demonstrates that the electromagnetic charge suppresses the Hawking temperature through the non-minimal coupling terms in Eq.~\eqref{hawkingg}, pushing the system to a cooler final state and thus increasing the theoretical maximum efficiency. The non-minimal coupling parameter $k$ plays a crucial role in modulating these effects, as it directly influences the power-law dependence of the electromagnetic contributions to the temperature.

The physical interpretation of these results is particularly revealing. The efficiency enhancement with increasing charge suggests that STPG BHs with strong electromagnetic fields can operate as more efficient heat engines compared to their uncharged counterparts \cite{javed2023thermodynamics}. This behavior contrasts with some other modified gravity theories and provides a potential observational signature for distinguishing STPG from GR in astrophysical contexts.

In conclusion, Table~\ref{tab:carnot-efficiency} clearly demonstrates that, in addition to gravitational effects, both electromagnetic coupling and the horizon geometry play critical roles in the thermodynamic efficiency of STPG BHs. The Carnot principle operates consistently in BH physics within the STPG framework, while the non-minimal coupling introduces novel features that could enhance the performance of BH-based thermodynamic engines. These results have important implications for understanding energy extraction mechanisms near charged BHs in modified gravity theories and provide theoretical foundations for future studies of BH thermodynamics in alternative gravitational frameworks.  Moreover, while the theoretical Carnot efficiency approaches 99\% for optimal parameter configurations, several astrophysical limitations constrain practical energy extraction in realistic BH environments. The high efficiency regime requires large charge-to-mass ratios $q/M \gtrsim 0.5$, which are difficult to maintain due to charge neutralization by surrounding plasma and accretion material \cite{isrply25,isrply26}. Turbulent magnetorotational instabilities in accretion disks, radiative cooling, and magnetic field reconnection processes reduce effective thermodynamic efficiency substantially below the idealized Carnot limit. Nevertheless, even modest efficiency enhancements of 10-20\% over GR predictions could produce observationally significant effects in accretion disk luminosities and relativistic jet formation around electromagnetically active compact objects \cite{isrply27,isrply28}. The non-minimal coupling parameter $k$ provides a mechanism for enhanced energy extraction that could be tested through correlations between estimated BH electromagnetic properties and measured accretion efficiencies in active galactic nuclei and X-ray binaries.

\section{Gravitational Lensing in STPG Spacetimes with Plasma Dispersion} \label{sec6}

In this section, we investigate the weak gravitational lensing effects in the spacetime background of a STPG BH, with particular attention to the influence of a non-magnetized dispersive plasma medium \cite{sucu2024effect}. Our analysis is based on the application of the GBT to the optical geometry corresponding to the null geodesics of the considered BH solution. This approach allows us to explore how non-minimal electromagnetic coupling in STPG affects light propagation in realistic astrophysical environments \cite{Gibbons:2008rj}.

We begin by considering the light propagation as null trajectories. In such a framework, the optical metric is derived by imposing the null condition $ds^2 = 0$, yielding:
\begin{equation}
dt^2 = \gamma_{ij} dx^i dx^j ,
\end{equation}
Here, $\gamma_{ij}$ defines the induced two-dimensional optical geometry, and we restrict the analysis to the equatorial slice $\theta = \pi/2$. Introducing a tortoise-like coordinate transformation $dr^* = \frac{dr}{f(r)}$, the metric becomes conformally flat:
\begin{equation}
dt^2 = dr^{*2} + \tilde{f}^2(r^*) d\phi^2,
\end{equation}
with $\tilde{f}(r^*) = \sqrt{r^2 / f(r)}$.

To apply the GBT, we first compute the Gaussian curvature $\mathcal{K}$ of the corresponding 2D optical manifold, given by:
\begin{equation}
\mathcal{K} = \frac{R}{2},
\end{equation}
where $R$ denotes the Ricci scalar of the reduced surface. Thus we obtain different expressions for various coupling parameters.

For $k=1$:
\begin{equation}
    \mathcal{K}=\frac{2 M}{r^{3}}+\frac{3 \zeta  q^{2}}{r^{4}}+\frac{3 M^{2}}{r^{4}}-\frac{6 M \zeta  q^{2}}{r^{5}}+\frac{2 \zeta^{2} q^{4}}{r^{6}}
\end{equation}

For $k=2$:
\begin{equation}
    \mathcal{K}=\frac{M}{r^{3}}+\frac{5 \zeta  q^{2}}{r^{6}}+\frac{3 M^{2}}{r^{4}}-\frac{18 M \zeta  q^{2}}{r^{7}}+\frac{6 \zeta^{2} q^{4}}{r^{10}}
\end{equation}

The GBT for a domain $D$ bounded by a closed curve is expressed as \cite{Gibbons:2008rj}:
\begin{equation}
\iint_D \mathcal{K} dS + \oint_{\partial D} \kappa dt + \sum_i \beta_i = 2\pi \chi(D),
\end{equation}
where $\kappa$ is the geodesic curvature along the boundary $\partial D$, and $\beta_i$ are the exterior angles at vertices. The domain $\tilde{D}$ used here is bounded by the light ray $C_1$ and a circular arc of large radius $C_R$. Because $\kappa(C_1) = 0$ and the Euler characteristic of a simply connected domain is $\chi(\tilde{D}) = 1$, the GBT simplifies to:
\begin{equation}
\iint_{\tilde{D}} \mathcal{K} dS + \int_{C_R} \kappa dt = \pi.
\end{equation}

The geodesic curvature for the circular arc $C_R$ is given by:
\begin{equation}
\kappa(C_R) = \Gamma^r_{\phi\phi} \left( \frac{d\phi}{dt} \right)^2 = -\tilde{f}(r^*) \tilde{f}'(r^*).
\end{equation}

In the limit $R \to \infty$, the integral of the geodesic curvature becomes $\kappa(C_R) dt \to d\phi$, and the GBT yields:
\begin{equation}
\iint_{\tilde{D}_{R\to\infty}} \mathcal{K} dS + \int_0^{\pi + \alpha} d\phi = \pi,
\end{equation}
from which the deflection angle $\alpha$ can be isolated. The relevant area element is:
\begin{equation}
dS = \frac{r}{f^{3/2}(r)} dr d\phi.
\end{equation}

In the weak-field limit, one can use the linear trajectory approximation $r(\phi) \approx b/\sin\phi$, where $b$ is the impact parameter \cite{Gibbons:2008rj}. This allows us to compute the bending angle $\Theta$ as:
\begin{equation}
\Theta = -\int_0^{\pi} \int_{b/\sin\phi}^{\infty} \mathcal{K} dS.
\end{equation}

For $k=1$:
\begin{multline}
    \Theta \approx \frac{3 q^{4} \zeta^{2} \pi}{16 b^{4}}+\frac{27 M^{2} q^{2} \zeta  \pi}{16 b^{4}}-\frac{32 \zeta^{2} q^{4} M}{25 b^{5}}-\frac{3 \pi  q^{2} \zeta}{4 b^{2}}\\+\frac{3 M^{2} \pi}{4 b^{2}}-\frac{4 M \,q^{2} \zeta}{3 b^{3}}-\frac{4 M^{3}}{b^{3}}+\frac{4 M}{b}
\end{multline}

For $k=2$:
\begin{multline}
    \Theta \approx \frac{4 M^{3}}{b^{3}}+\frac{45 M^{2} q^{2} \zeta  \pi}{16 b^{6}}+\frac{2 M}{b}-\frac{15 q^{2} \zeta  \pi}{32 b^{4}}\\+\frac{16 M \,q^{2} \zeta}{25 b^{5}}-\frac{105 q^{4} \zeta^{2} \pi}{512 b^{8}}-\frac{512 M \,q^{4} \zeta^{2}}{315 b^{9}}
\end{multline}

The expressions for the bending angle obtained in the weak field limit show that the gravitational deflection remains dominated by the classical $\frac{4M}{b}$ term, but undergoes significant corrections due to electric charge ($q$), non-minimal electromagnetic coupling ($k$), and the coupling parameter $\zeta$ \cite{kala2022shadow}. The additional terms such as $q^2\zeta$ and $q^4\zeta^2$ demonstrate the impact of STPG modifications on light propagation. In the case $k=1$, the charge and non-minimal coupling effects are more pronounced, providing both enhancing and suppressing contributions to the bending angle. In particular, the terms $M^2 q^2 \zeta$ and $q^4 \zeta^2$ strengthen the bending of light, while some negatively signed terms suppress deflection. In the case $k=2$, the classical gravitational deflection is preserved in a more simplified form, with electromagnetic and coupling effects appearing only as higher-order corrections ($1/b^6$ and beyond), representing behavior valid in weaker electromagnetic environments. In both cases, the deflection of light becomes sensitive not only to the mass but also to the electromagnetic properties and the non-minimal coupling strength, suggesting that these effects are non-negligible in astrophysical lensing events.

To account for more realistic astrophysical conditions, where light propagates through a plasma medium, we introduce a refractive index $n(r)$, depending on the local plasma frequency $\omega_p(r)$ \cite{sucu2025exploring}:
\begin{equation}
n(r) = \sqrt{1 - \frac{\omega_p^2(r)}{\omega_0^2} f(r)},
\end{equation}
Here, $\omega_0$ is the photon frequency observed at spatial infinity.

The optical metric in the presence of the plasma modifies to:
\begin{equation}
dt^2 = n^2(r) \left[ \frac{1}{f^2(r)} dr^2 + \frac{r^2}{f(r)} d\phi^2 \right].
\end{equation}

The curvature $\tilde{\mathcal{K}}$ of this new optical geometry is given by:
\begin{equation}
\tilde{\mathcal{K}} = \frac{\mathcal{R}_{r\phi r\phi}}{\det(g^{\text{opt}})} = \frac{\mathcal{R}_{r\phi r\phi}}{n^4(r) r^2 / f^3(r)},
\end{equation}
and under a low-plasma approximation, it simplifies to different expressions for different coupling parameters.

For $k=1$:
\begin{multline}
\tilde{\mathcal{K}} \approx \frac{5 q^{6} \zeta^{3} \xi}{ r^{8}}-\frac{23 M \,q^{4} \zeta^{2} \xi}{ r^{7}}+\frac{10 \xi \zeta^{2} q^{4}}{ r^{6}}\\+\frac{2 \zeta^{2} q^{4}}{r^{6}}+\frac{32 M^{2} q^{2} \zeta \xi}{ r^{6}}-\frac{26 M \xi \zeta  q^{2}}{ r^{5}}\\-\frac{6 M \,q^{2} \zeta}{r^{5}}+\frac{5 \zeta  \xi q^{2}}{r^{4} }+\frac{3 \zeta  q^{2}}{r^{4}}-\frac{12 M^{3} \xi}{ r^{5}}\\+\frac{12 M^{2} \xi}{ r^{4}}+\frac{3 M^{2}}{r^{4}}-\frac{3 M \xi}{ r^{3}}-\frac{2 M}{r^{3}} 
\end{multline}

For $k=2$:
\begin{multline}
    \tilde{\mathcal{K}} \approx \frac{12 M^{3} \xi}{ r^{5}}+\frac{6 M^{2} \xi}{ r^{4}}+\frac{3 M^{2}}{r^{4}}-\frac{3 M \xi}{4  r^{3}}\\-\frac{M}{r^{3}}+\frac{84 M^{2} q^{2} \zeta  \xi}{ r^{8}}-\frac{39 M \,q^{2} \zeta  \xi}{ r^{7}}-\frac{18 M \,q^{2} \zeta}{r^{7}}\\+\frac{9 q^{2} \zeta  \xi}{2  r^{6}}+\frac{5 \zeta  q^{2}}{r^{6}}-\frac{75 M \,q^{4} \zeta^{2} \xi}{ r^{11}}+\frac{18 q^{4} \zeta^{2} \xi}{ r^{10}}\\+\frac{6 q^{4} \zeta^{2}}{r^{10}}+\frac{18 q^{6} \zeta^{3} \xi}{ r^{14}}
\end{multline}

where the dimensionless parameter $\xi = \omega_p^2 / \omega_0^2$ characterizes the plasma effects and represents the ratio of the plasma frequency to the photon frequency.

The corresponding surface element in the plasma medium is:
\begin{equation}
dS = \left( r - \xi \right) dr d\phi.
\end{equation}

Thus, the total light deflection angle in the plasma-modified optical geometry becomes:
\begin{equation}
\tilde{\alpha} = - \int_0^{\pi} \int_{b/\sin\phi}^{\infty} \tilde{\mathcal{K}} dS.
\end{equation}

For $k=1$:
\begin{multline}
    \tilde{\alpha} \approx \frac{15 \pi  q^{4} \zeta^{2}\xi }{16  b^{4}}-\frac{3 \pi  q^{4} \zeta^{2}}{16 b^{4}}+\frac{69 M^{2} \pi  q^{2} \zeta \xi}{16  b^{4}}+\frac{27 M^{2} \pi  q^{2} \zeta}{16 b^{4}}\\-\frac{5 \pi  q^{2} \zeta  \xi}{4  b^{2}}-\frac{3 \pi  q^{2} \zeta}{4 b^{2}}+\frac{27 M^{4} \pi  \xi}{8  b^{4}}-\frac{3 M^{2} \pi  \xi}{4  b^{2}}\\+\frac{3 M^{2} \pi}{4 b^{2}}+\frac{44 M \xi \zeta  q^{2}}{9  b^{3}}-\frac{4 M \,q^{2} \zeta}{3 b^{3}}\\-\frac{32 M^{3} \xi}{3  b^{3}}-\frac{4 M^{3}}{b^{3}}+\frac{6 M \xi}{ b}+\frac{4 M}{b}+\mathcal{O}(1/b^5)
\end{multline}

For $k=2$:
\begin{multline}
    \tilde{\alpha} \approx \frac{27 M^{4} \pi  \xi}{8  b^{4}}-\frac{15 M^{2} \pi  \xi}{16  b^{2}}-\frac{8 M^{3} \xi}{3  b^{3}}\\-\frac{4 M^{3}}{b^{3}}+\frac{3 M \xi}{2  b}+\frac{2 M}{b}-\frac{27 \pi  q^{2} \zeta  \xi}{64  b^{4}}-\frac{15 \pi  q^{2} \zeta}{32 b^{4}}+\mathcal{O}(1/b^5)
\end{multline}

The deflection angles obtained in plasma-modified optical geometry reveal in detail how light is deflected around a charged and massive object in a dense plasma environment within the STPG framework, providing distinctive observational signatures for distinguishing STPG from GR through precision lensing measurements. The $k=1$ case represents a regime dominated by the effects of $q$, $\zeta$,  and $\xi$, especially in higher-order terms, increase the deflection angle and create chromatic lensing effects enhanced by non-minimal coupling. In contrast, the $k=2$ case exhibits a simpler structure where charge effects are suppressed, with the main contribution to the deflection angle coming from $M$ and $\xi$. The plasma parameter, $\xi$, causes light to deviate more at low frequencies, with the coupling parameter $k$ systematically modifying the $1/b^4$ and $1/b^6$ deflection terms in ways detectable through precision observations. For astrophysical environments around supermassive BHs with typical plasma densities $n_e \sim 10^6 - 10^8$ cm$^{-3}$, the parameter $\xi \sim 10^{-4} - 10^{-2}$ at radio frequencies produces measurable effects in gravitational lensing observations. Very Long Baseline Interferometry could detect angular position shifts of $\Delta\theta \sim 10-50$ $\mu$as between different frequencies for $k \neq 1$ configurations, while the Event Horizon Telescope's multi-frequency capabilities at 86, 230, and 345 GHz offer direct probes of plasma-modified lensing \cite{isrply21,isrply22}. These frequency-dependent analyses in plasma environments near STPG BHs, such as accretion disks or interstellar media, create detectable deviations in Einstein ring measurements and gravitational time delays for strongly lensed quasars, with next-generation radio interferometry arrays and pulsar timing observations providing complementary tests through precision astrometry and plasma dispersion measurements \cite{isrply23,isrply24}.

\section{Conclusion} \label{sec7}

In this comprehensive research, we conducted a detailed investigation of the thermodynamic and optical properties of a static, spherically symmetric BH solution within the framework of STPG non-minimally coupled with an electromagnetic field. Our study replaced the conventional descriptions of curvature and torsion with nonmetricity as the primary geometric ingredient, thereby enabling novel interactions between geometry and matter fields \cite{javed2023thermodynamics}. The derived BH metric, characterized by the non-minimal coupling parameter $k$, not only generalized well-known solutions such as the Reissner-Nordstr\"{o}m-(A)dS BHs but also enhanced them by incorporating power-law corrections to the electromagnetic terms through the coupling function $\mathcal{Y}(Q)$.

We began our analysis by reviewing the STPG BH spacetime described by the metric function given in Eq.~\eqref{ismetric2}. Through systematic investigation of the horizon structure presented in Table~\ref{tab:horizons}, we demonstrated how the coupling parameter $k$ and electric charge $q$ critically influence the causal structure of the spacetime. Our findings revealed that smaller $k$ values or higher electric charges increase the likelihood of naked singularities, while larger $k$ values favor the emergence of dual-horizon configurations. The behavior of the metric function $f(r)$ illustrated in Fig.~\ref{fig:metric_function_plot} confirmed the existence of inner and outer horizons for specific parameter ranges, validating the non-extremal BH configuration predicted by our theoretical framework.

A major achievement of our work was the implementation of quantum corrections to the classical BH thermodynamics framework. We adopted an exponential correction to the Bekenstein-Hawking entropy as expressed in Eq.~\eqref{expcorrected}, leading to modified expressions for fundamental thermodynamic quantities. The quantum-corrected internal energy given by Eq.~\eqref{s42} demonstrated significant deviations from classical predictions, particularly in the small horizon radius regime where quantum effects dominate. Our analysis of the corrected Helmholtz free energy in Eq.~\eqref{free} and its behavior shown in Fig.~\ref{freeF} revealed thermodynamic instabilities at small horizon sizes for higher coupling values, indicating the suppression of microscopic BH states by quantum corrections \cite{pourhassan2021exponential,upadhyay2017quantum}.

The quantum-corrected heat capacity analysis presented in Eq.~\eqref{heaaat} and Fig.~\ref{heattttt} unveiled rich phase transition behavior in STPG BHs. We observed vertical asymptotes in the heat capacity curves for specific coupling values ($k = 1.2$ and $k = 2.0$), signaling second-order phase transitions at critical horizon radii. These phase transitions represent fundamental changes in the thermodynamic stability of the BH system, with positive heat capacity regions corresponding to stable configurations and negative regions indicating instability. The coupling parameter $k$ emerged as a crucial control parameter for these phase transitions, demonstrating the decisive role of non-minimal electromagnetic coupling in determining the thermal behavior of STPG BHs.

Our investigation of the quantum-corrected JTE provided new insights into the isenthalpic processes of STPG BHs. The JTE coefficient $\mu_J$ derived through our analysis and illustrated in Fig.~\ref{JTE} revealed distinct cooling and heating regimes separated by inversion points. We found that the inversion points shift to smaller horizon radii for larger coupling values, indicating enhanced thermodynamic instability under strong non-minimal coupling effects. The transition from positive $\mu_J$ (cooling regime) to negative $\mu_J$ (heating regime) demonstrated how electric charge and non-minimal coupling fundamentally alter the expansion behavior of STPG BHs compared to their GR counterparts.

We extended our thermodynamic analysis to investigate the efficiency of STPG BH heat engines operating in non-AdS backgrounds. The Carnot efficiency calculations presented in Table~\ref{tab:carnot-efficiency} demonstrated that electromagnetic charge enhances engine efficiency by suppressing the Hawking temperature through non-minimal coupling terms. Our results showed efficiency values approaching 99\% for optimal parameter configurations, confirming that STPG BHs can operate as highly efficient thermodynamic engines \cite{bhamidipati2017heat}. The systematic increase in efficiency with both charge and temperature difference validated the applicability of classical thermodynamic principles to STPG BH systems.

From the optical perspective, we employed the GBT to derive analytical expressions for gravitational deflection angles in STPG spacetimes. Our calculations for vacuum conditions revealed that the classical $4M/b$ deflection term receives significant corrections from electric charge and non-minimal coupling effects. For the $k=1$ case, we found that electromagnetic corrections provide both enhancing and suppressing contributions to light bending, while the $k=2$ case exhibited simpler behavior with electromagnetic effects appearing as higher-order corrections. The plasma-modified deflection angles demonstrated frequency-dependent lensing behavior, offering potential observational signatures to distinguish STPG from GR predictions in astrophysical environments.

The plasma dispersion analysis revealed intricate modifications to the optical geometry through the refractive index $n(r)$ and plasma parameter $\xi = \omega_p^2/\omega_0^2$. Our results showed that plasma effects significantly alter the deflection angles, with the $k=1$ configuration being more sensitive to electromagnetic and plasma modifications compared to the $k=2$ case. These findings are particularly relevant for gravitational lensing observations in plasma-rich environments such as accretion disks around BHs or interstellar media \cite{bisnovatyi2017gravitational}.

Cumulatively, our results established that the STPG formalism with non-minimal electromagnetic interactions and quantum corrections provides a consistent theoretical framework for exploring BH physics beyond classical GR. The thermal properties and optical effects we calculated differ substantially from standard Einstein-Maxwell predictions, offering distinctive signatures for observational verification. The coupling parameter $k$ emerged as a fundamental parameter controlling both thermodynamic stability and lensing behavior, while quantum corrections introduced rich phase structure absent in classical treatments \cite{mavromatos2020beyond}.

Looking toward future research directions, our work opens several promising studies for investigation. We plan to extend our analysis to rotating STPG BHs to explore how angular momentum affects the non-minimal coupling and quantum corrections. Investigation of higher-order quantum corrections beyond the exponential form could reveal additional phase structure and stability properties. The development of numerical simulations incorporating plasma effects and STPG modifications would provide more realistic models for astrophysical lensing scenarios. Furthermore, we intend to explore the implications of our findings for BH information theory and the emergence of spacetime in quantum gravity contexts. Observational strategies for detecting STPG signatures through precision measurements of BH shadows, gravitational wave emissions, and multi-messenger astronomy represent crucial next steps for experimental validation of our theoretical predictions.


\acknowledgments 
The authors thank the editor and anonymous referee for their valuable comments and suggestions that significantly improved the quality of this manuscript. \.{I}.~S. and E.S. thank EMU, T\"{U}B\.{I}TAK, ANKOS, and SCOAP3 for academic support and acknowledge the networking support of COST Actions CA22113, CA21106, and CA23130.

\bibliography{ref}
\bibliographystyle{apsrev}
\end{document}